\documentclass[%
 reprint,
superscriptaddress,
 amsmath,amssymb,amsfonts,
 aps,
pra,
 floatfix,
 longbibliography,nobibnotes
]{revtex4-2}

\usepackage[utf8]{inputenc}

\usepackage[pdftex]{graphicx} \graphicspath{{}}
\usepackage{xcolor}
\usepackage[colorlinks=true,linkcolor=blue,urlcolor=cyan,citecolor=blue]{hyperref}

\usepackage{amsmath} \usepackage{amsfonts} \usepackage{amssymb}
\usepackage{mathtools}
\DeclarePairedDelimiter\bra{\langle}{\rvert}
\DeclarePairedDelimiter\ket{\lvert}{\rangle}
\DeclarePairedDelimiterX\braket[2]{\langle}{\rangle}{#1 \delimsize\vert #2}
\DeclarePairedDelimiterX\expval[3]{\langle}{\rangle}{#1 \delimsize\vert #2  \delimsize\vert #3}
\DeclarePairedDelimiterX\proj[2]{\delimsize\vert#1\rangle}{\langle#2\delimsize\vert}{ }

\usepackage[final]{microtype}

\usepackage{booktabs}

\begin{document}

\title{Scalable spin-nematic squeezing in multi-level dipole-interacting Rydberg atom arrays}
\author{Sakshi Bahamnia}
\email{sakshi.bahamnia@ou.edu}
\affiliation{Homer L. Dodge Department of Physics and Astronomy, The University of Oklahoma, Norman, Oklahoma 73019, USA}
\affiliation{Center for Quantum Research and Technology, The University of Oklahoma, Norman, Oklahoma 73019, USA}
\author{Thomas Bilitewski}
\email{thomas.bilitewski@okstate.edu}
\affiliation{Department of Physics, Oklahoma State University, Stillwater, Oklahoma 74078, USA} 

\date{\today}

\begin{abstract}

We study the dynamical generation of metrologically useful entanglement in arrays of three-level (spin-1) systems coupled by spin-exchange interactions, as realized by dipole-interacting Rydberg atoms in optical tweezers. %
We quantify the metrological gain by a generalized squeezing parameter optimized over all linear $\mathrm{SU}(3)$ phase generators and readout observables. For equal exchange couplings on the two transitions, the dynamics reduces to one-axis twisting (OAT) of an effective two-level system. Away from this point, the full three-level structure provides a genuine advantage. For all-to-all couplings, the optimal squeezing at $N=4000$ exceeds the OAT value by a factor of 3.6, with a fitted scaling exponent of $0.93$ compared with $0.69$ for OAT over $100\le N\le4000$. %
For two-dimensional dipolar arrays, exact and tensor-network simulations up to $N=49$ show that the optimal three-level protocol outperforms the effective spin-1/2 dynamics, with a gain that increases with system size, and semiclassical simulations show that squeezing continues to grow up to $N=10^4$. %
Crucially, our results demonstrate a real advantage of utilizing the full three-level $\mathrm{SU}(3)$ structure of the spin-1 system over $\mathrm{SU}(2)$ type one-axis twisting dynamics. %
Our results extend interaction-driven scalable squeezing beyond the well-studied qubit setting to spin-exchange-coupled qudits with dipolar interactions, and can be realized in current Rydberg tweezer experiments.
\end{abstract}
\maketitle
%
%
{\it Introduction.---}%
Entangled quantum states can enhance the precision of measurements beyond the standard quantum limit (SQL), a prospect that has motivated intense efforts in quantum metrology~\cite{Giovannetti2011,Pezze2018,RevModPhys.89.035002,Montenegro_2025}. Spin squeezing, the redistribution of quantum projection noise from one spin quadrature to another, is among the most established routes to such metrological gain~\cite{Kitagawa1993,Wineland1992,Wineland1994,Ma2011}, and is directly linked to multipartite entanglement~\cite{Sorensen2001,Pezze2018}. The paradigmatic one-axis twisting (OAT) model shows that collective all-to-all interactions can generate spin-squeezed states whose sensitivity improves algebraically with the number of particles $N$~\cite{Kitagawa1993}, and the closely related two-axis countertwisting (TAT) model can in principle reach the Heisenberg limit $\xi^2\propto 1/N$~\cite{Kitagawa1993}. Squeezing via all-to-all interactions has been demonstrated experimentally in a variety of platforms, including atoms in optical cavities~\cite{Leroux2010,Hosten2016,PedrozoPenafiel2020}, trapped-ion crystals~\cite{RevModPhys.93.025001,Bohnet2016,Franke2023}, and Bose--Einstein condensates~\cite{Esteve2008,Lucke2011,Gross2011,Riedel2010,Hamley2012}.

A central question is whether these mechanisms extend to spatially extended systems with only finite-range interactions, where the simple collective picture breaks down. A series of theoretical works has demonstrated that scalable squeezing can also arise from short-range and power-law decaying interactions in two and three spatial dimensions \cite{Rey2008,Foss-Feig2016,Perlin2020,RoscildePRA2021,Comparin2022a,Comparin2022b,Roscilde2023,Wellnitz2024,Roscilde2024,Roscilde2024b,Bilitewski2021}, provided the underlying Hamiltonian supports easy-plane ferromagnetic order at finite temperature~\cite{Block2024,Kaplan-Lipkin2025}. Beyond one-axis twisting, two-mode squeezing via pair creation~\cite{BilitewskiRey2023,Duha2024,Duha_2025,Duha_2026,Sundar2023} and countertwisting with gap protection~\cite{Koyluoglu2025,bahamnia2026gapprotectedheisenberglimitedsqueezinglocally} have been proposed as routes to scalable entanglement generation in power-law interacting systems, and two-axis twisting has been realized with Floquet-engineered polar molecules \cite{Miller2024} and in optical cavities \cite{Luo2025}.

These theoretical advances have stimulated a wave of experimental demonstrations of scalable squeezing with finite-range interactions across a variety of platforms. Arrays of neutral atoms individually trapped in optical tweezers and coupled via Rydberg-state interactions have emerged as a leading platform~\cite{Saffman2010,Browaeys2020,Henriet2020quantumcomputing,Morgado2021,Kaufman2021}, enabling landmark demonstrations of scalable spin squeezing with dipolar interactions in two-dimensional geometries~\cite{Bornet2023}, closely connected to the realization of continuous symmetry breaking~\cite{Chen2023,Sbierski2024}. Further demonstrations include Rydberg-mediated squeezing in optical clocks~\cite{Eckner2023} and Rydberg-dressed atomic ensembles~\cite{Hines2023}, alongside squeezing with finite-range interactions in trapped-ion chains~\cite{Franke2023} and lattice gases with contact interactions~\cite{Lee2025}.  In parallel, magnetic atoms~\cite{Chomaz2023} in optical lattices~\cite{Lepoutre2019,Roscilde2024,Douglas2025}, ultracold polar molecules~\cite{Baranov2012,Bohn2017,Moses2017,Bilitewski2021,Wellnitz2024,Christakis2023,Carroll2025,holland2026}, and solid-state spin defects~\cite{Wu2025} have all demonstrated, or are predicted to support, scalable entanglement generation via dipolar spin-exchange interactions.

This prior work has focused predominantly on effective two-level, spin-1/2 systems. Yet Rydberg atoms naturally offer a ladder of states with alternating parity that can encode multi-level, qudit degrees of freedom~\cite{Zache2022,Lienhard2020,Zhang2024,Mogerle2025,Qiao2025}, and external microwave and optical fields provide local control over the resulting Hamiltonian~\cite{Bornet2024LocalControl}. Moving beyond qubits to higher-dimensional internal spaces opens qualitatively new physics. For spin-1 systems, the relevant operator algebra extends from SU(2) to SU(3), admitting both spin and nematic (quadrupole) order parameters and correspondingly richer classes of squeezing~\cite{Yukawa2013,Yukawa_2016,StamperKurn2013}. Spin-nematic squeezing has been studied extensively in spinor Bose--Einstein condensates, where collisional spin-mixing dynamics generate entanglement between spin and quadrupole quadratures~\cite{Hamley2012,Mao2023,Xin2023}, and two-mode-squeezing Hamiltonians have been proposed for multi-level dipoles coupled to optical cavities~\cite{Sundar2023}. However, the question whether spatially extended arrays of dipole-interacting spin-1 particles, without cavity-mediated all-to-all interactions, can produce scalable squeezing that exploits the full SU(3) structure, and whether this yields a genuine advantage over (effective) two-level dynamics, has remained largely unexplored.

In this work, we address this question by studying a three-level spin-1 system. %
We first consider the all-to-all interacting limit to elucidate the entanglement-generation mechanism, using a generalized squeezing parameter that quantifies the metrological gain for the optimal linear $\mathrm{SU}(3)$ generator and readout operations. We demonstrate a metrological advantage of the three-level dynamics over effective two-level OAT, and a trade-off between the achievable squeezing and the squeezing time. We then show that this advantage persists for dipolar interactions in two-dimensional arrays, and provide explicit Rydberg level schemes realizing the required interactions. %
These results establish scalable entanglement generation in multi-level Rydberg systems and constitute a first step toward harnessing the richer internal structure of qudits for quantum-enhanced metrology.

{\it Model.---}%
\begin{figure}[t]
    \centering  
    \includegraphics[width=1.0\linewidth]{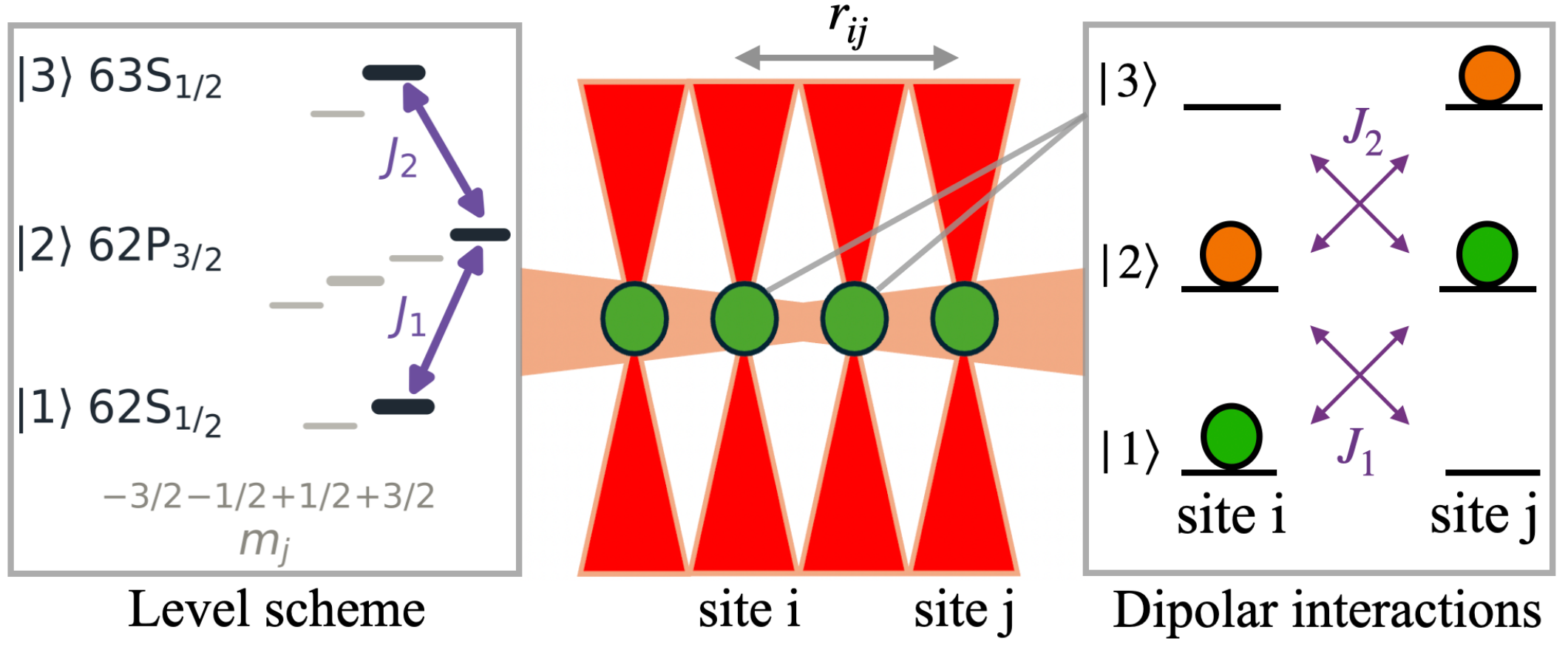}
    \caption{Model overview. Rydberg states with alternating parity define a three-level system labeled $\{ \ket{1}, \ket{2}, \ket{3} \}$. Microwaves $\Omega_{1,2}$ drive transitions between adjacent levels, and dipolar interactions $J_{ij}^{1(2)}$ induce pairwise spin exchange within the 1--2 and 2--3 sublevels between atoms at sites $i$ and $j$ with strength $\sim r_{ij}^{-3}$.}
    \label{fig:fig1Threelevelatomicarray}
\end{figure}
We consider an array of $N$ atoms trapped individually in optical tweezers, illustrated in
Fig.~\ref{fig:fig1Threelevelatomicarray}. We consider a two-dimensional square lattice with spacing $a$; sites are labeled by $i,j$, and the distance between atoms at sites $i$ and $j$ by $r_{ij}$. %
Each atom can occupy any rung of a ladder of Rydberg states denoted as $|\nu\rangle$ with
$\nu = 1, 2, \ldots$, where the levels alternate in parity. Each $\ket{\nu}$ is a single
Zeeman sublevel $(n_\nu,\ell_\nu,j_\nu,m_{j,\nu})$ of a Rydberg fine-structure manifold, and consecutive
rungs are connected by an electric-dipole-allowed transition ($\Delta\ell=\pm1$), so that the
ladder is a sequence of alternating-parity manifolds such as $nS_{1/2}$--$nP_{3/2}$--$nD_{5/2}$ or
$nP_{1/2}$--$nS_{1/2}$--$nP_{3/2}$, with explicit level schemes provided in Appendix~\ref{app:candidates}.
A static magnetic field defines the quantization axis and Zeeman-splits each manifold, both isolating the chosen sublevels and detuning competing exchange processes that would otherwise transfer population out of them (Appendix~\ref{app:neglected}). 
A resonant microwave field $\Omega_{\nu}$ couples adjacent Rydberg states, inducing transitions
between levels $\nu$ and $\nu+1$, while atoms at different sites interact predominantly via
dipole-dipole interactions with strength $J_{ij}^{\nu}$ through the exchange process
$|\nu \rangle _{i} |\nu+1 \rangle _{j} \leftrightarrow |\nu+1 \rangle _{i} |\nu\rangle _{j}$.

The general many-body Hamiltonian is given by
\begin{equation}
\begin{aligned}
    \hat{H} &= \sum_{\nu} \sum_{i \neq j} J^\nu_{ij} \left(\vert \nu+1 \rangle \langle \nu \vert\right)_i\left(\vert \nu \rangle \langle \nu+1 \vert\right)_j \\
    &+ \sum_{\nu,i}\frac{\Omega_\nu}{2}\Big[ \left(\vert \nu \rangle \langle \nu+1 \vert\right)_i +\left(\vert \nu+1 \rangle \langle \nu \vert\right)_i\Big] ,
    \end{aligned} \label{eq:multilevelHamiltonian}
\end{equation}
with $J_{ij}^\nu = J_\nu\,[1-3\cos^2\theta_{ij}]/r_{ij}^3$, where $r_{ij}$ is measured in units of $a$ and $\theta_{ij}$ is the angle between $\mathbf r_{ij}$ and the quantization axis. We take the magnetic field perpendicular to the array, so that $\theta_{ij}=\pi/2$ for all pairs and $J_\nu$ is the nearest-neighbor coupling. The coupling $J_\nu/2\pi=\kappa\,\mathcal{A}_\nu\,|M_\nu|^2/a^3$ [Eq.~\eqref{eq:Ji}] is set by the dimensionless transition dipole matrix element $M_\nu$ of the $\ket{\nu}\!\leftrightarrow\!\ket{\nu+1}$ transition, and its sign by the polarization of the transition: a $\pi$ transition ($\Delta m_j=0$) has $\mathcal{A}_\nu=+1$, whereas a $\sigma$ transition ($|\Delta m_j|=1$) has $\mathcal{A}_\nu=-1/2$ (Appendix~\ref{app:dipolar_exchange}).

Within this general multi-level framework, we restrict our focus to the simplest case beyond a qubit, a three-level system (qutrit), labeling three consecutive states $\nu,\nu+1,\nu+2$ as $1,2,3$. %
We define the operators $\hat{T}^{\mu \nu}_i = (|\mu \rangle \langle \nu|)_i$, representing populations when $\mu = \nu$ and transition operators when $\mu \neq \nu$. The Hamiltonian in Eq.~\eqref{eq:multilevelHamiltonian} then takes the form $\hat{H} = \hat{H}_{\Omega} + \hat{H}_{\text{int}}$ with
\begin{equation}
    \begin{aligned}    
    \hat{H}_{\Omega} = &\sum_{i} \bigg[ \frac{\Omega_{1}}{2} \left( \hat{T}_{i}^{12} + \hat{T}_{i}^{21} \right)  + \frac{\Omega_{2}}{2}  \left( \hat{T}_{i}^{23} + \hat{T}_{i}^{32} \right) \bigg],\\
    \hat{H}_{\text{int}} = & \sum_{i \neq j} \bigg[ J_{ij}^{1} \hat{T}^{21}_{i} \hat{T}^{12}_{j} + J^{2}_{ij} \hat{T}^{32}_{i} \hat{T}^{23}_{j} \bigg],
    \end{aligned} 
    \label{eq:ThreelevelFullHamiltonian}
\end{equation}
where $\Omega_1$ ($\Omega_2$) couples the states $\ket{1}$ and $\ket{2}$ ($\ket{2}$ and $\ket{3}$), and $J_{1}$ ($J_{2}$) is the dipolar interaction strength for the sublevels $\{\ket{1}, \ket{2}\}$ ($\{\ket{2}, \ket{3}\}$). We define the ratio $r = J_{2}/ J_{1}$ which measures the strength of the 2–3 exchange relative to the 1–2 exchange. Focusing without loss of generality on $|r| \le 1$, we set $J \equiv |J_1| = \max(|J_1|, |J_2|)$ as the reference energy scale. Since the dynamics of the squeezing parameter considered below is invariant under $\hat H\to-\hat H$, the overall sign of $J_1$ is irrelevant. We consider power-law couplings $J^\nu_{ij}=J_\nu/r_{ij}^{\alpha}$, with $\alpha=0$ (all-to-all, $J^\nu_{ij}=J_\nu$) and $\alpha=3$ (dipolar).

Restricting the dynamics to the three chosen levels requires that all other Rydberg states, as well as processes beyond the retained interactions, be energetically well separated or sufficiently weak. In Appendix~\ref{app:candidates} we substantiate this for the explicit level schemes.

After preparing a product initial state with the single-particle drives, we switch off the drives ($\Omega_\nu=0$), let the state evolve solely under $\hat H_{\mathrm{int}}$, and analyze the dynamical generation of entanglement in the form of spin squeezing. 

To characterize spin squeezing in the full $\mathrm{SU}(3)$ space, we utilize a generalized squeezing parameter $\xi^2_{\mathrm{opt}}$ \cite{bahamnia2026gapprotectedheisenberglimitedsqueezinglocally}, following the methodology of Ref.~\cite{Gessner2019} specialized to linear operators in SU(3). Specifically, we consider a metrological protocol of the following form: a phase imprinted onto the state by a linear generator $\hat H_{\mathbf n}=\mathbf n\cdot\hat{\boldsymbol\Lambda}$ and subsequently read out from a linear observable $\hat M_{\mathbf m}=\mathbf m\cdot\hat{\boldsymbol\Lambda}$, both expressed in terms of the full basis of collective Gell-Mann operators $\hat{\boldsymbol{\Lambda}}$, and define the squeezing parameter $\xi^2_{\mathrm{opt}}$ as the ratio of the best sensitivity achievable with a separable state to the sensitivity of the given state, optimized over generator and readout operators (see Appendix~\ref{app:xiopt}), so that $1/\xi_{\rm{opt}}^2$ is the metrological gain. We note that this reduces to the familiar Wineland squeezing parameter if restricting the generator and read-out operations to a SU(2) subset. It is also, by construction, an entanglement witness, since any value below 1 cannot be achieved by a separable state.

We obtain results for all-to-all interactions from exact calculations in the permutationally symmetric subspace, and for dipolar interactions from exact diagonalization (ED) for $N\le16$, matrix-product-state (MPS) simulations for $N\le49$, and the generalized discrete truncated Wigner approximation (dTWA)~\cite{Schachenmayer2015,Zhu_NewJournalofPhysics_21_2019} for larger systems.  A numerical benchmark comparison for our system between exact methods and dTWA is provided in Appendix~\ref{app:benchmarks}.

{\it Results.---}
We begin the discussion by considering all-to-all interactions, i.e. $J^{\nu}_{ij} = J_{\nu}$, which, although not directly realized by dipolar interactions, isolates the squeezing mechanism and establishes the metrological advantage of full three-level squeezing over two-level OAT dynamics  before we turn to demonstrating the effectiveness of these mechanisms for the case of dipolar interactions.

For symmetric interactions ($r=1$), the interaction Hamiltonian is $
    \hat H_{\text{int}}
    =
    \sum_{i\neq j}J_{ij}
    \left(
    \hat{T}^{21}_{i}\hat{T}^{12}_{j}
    +
    \hat{T}^{32}_{i}\hat{T}^{23}_{j}
    \right),
$
which, upon transforming to the basis $\{\ket{b},\ket{d},\ket{2}\}$ with bright and dark states defined as $\ket{b}=(\ket{1}+\ket{3})/\sqrt{2}$ and $\ket{d}=(\ket{1}-\ket{3})/\sqrt{2}$, takes the form
\begin{equation}
    \hat{H}= \sum_{i\neq j}J_{ij} \left(\hat{b}^{+}_{i}\hat{b}^{-}_{j}+\hat{d}^{+}_{i}\hat{d}^{-}_{j}\right),
      \label{Eq:BrightDarkHamiltonian}
\end{equation}
where $\hat{b}^{+}_{i}=(\ket{b}\!\bra{2})_i$, $\hat{b}^{-}_{i}=(\hat{b}^{+}_{i})^{\dagger}$, $\hat{b}^{z}_{i}=\tfrac12(\ket{b}\!\bra{b}-\ket{2}\!\bra{2})_i$ and $\hat{d}^{a}_{i}$ defined analogously. 

This form reveals two SU(2) manifolds, which are dynamically decoupled for $r=1$, a bright manifold with generators 
$\hat{\boldsymbol B}
=\{\tfrac12\hat{S}_{x},\,\tfrac12\hat{Q}_{yz},\,\tfrac14(\hat{D}_{xy}+ \sqrt{3}\hat{Y})\}$ and a dark manifold with generators $\hat{\boldsymbol D}=\{\tfrac12\hat{Q}_{xz},\,\tfrac12\hat{S}_{y},\,\tfrac14(\sqrt{3} \hat{Y}-\hat{D}_{xy})\}$, where the quadrupolar operators are defined in Appendix~\ref{app:operators}.  The generators of each manifold obey $\mathfrak{su}(2)$ commutation relations $[\hat B^a,\hat B^b]=i\epsilon_{abc}\hat B^c$ and $[\hat D^a,\hat D^b]=i\epsilon_{abc}\hat D^c$ (Appendix~\ref{app:bright_dark}). The bright manifold is spanned by $\{\ket{b}, \ket{2}\}$ effectively reducing dynamics to a two-level system, and analogously for the dark manifold, spanned by $\{\ket{d}, \ket{2}\}$. %
The Hamiltonian in Eq.~\eqref{Eq:BrightDarkHamiltonian} conserves the total number of bright and dark excitations $\hat N_B=\sum_i(\ket{b}\!\bra{b})_i$ and $\hat N_D=\sum_i(\ket{d}\!\bra{d})_i$ independently, ensuring that the dynamics remain decoupled within the corresponding manifolds. The bright-sector Hamiltonian can be written as $\hat{H}=\sum_{i\neq j} J_{ij}\, \hat{b}^{+}_{i}\hat{b}^{-}_{j}=\sum_{i\neq j} J_{ij}(\hat{\boldsymbol{b}}_{i} \cdot\hat{\boldsymbol{b}}_{j}-\hat{b}^{z}_{i}\hat{b}^{z}_{j})$, which, in the collective all-to-all limit, reduces to the familiar one-axis twisting (OAT) form $\hat{H}\propto J(\hat{B}^{z})^{2}$ up to constants of motion. An initial state $\ket{\psi}= \ket{b_x}^{\otimes N}$  with $\ket{b_x}=(\ket{b}+\ket{2})/\sqrt2$ therefore generates squeezing within the bright manifold, whereas $\ket{\psi}= \ket{d_x}^{\otimes N}$ produces analogous squeezing within the dark manifold. In both cases, the squeezing quadratures lie in the $y$--$z$ plane of the respective Bloch spheres, and the resulting squeezing is identified as type-2 spin-nematic squeezing~\cite{Yukawa2013,Hamley2012}. 
In the all-to-all limit the interaction takes the compact form $\hat H_{\mathrm{int}}=J_1\hat N_1\hat N_2+J_2\hat N_2\hat N_3=J\,\hat N_2(\hat N_1+r\hat N_3)$ within the permutationally symmetric subspace, with $\hat N_\nu=\sum_i(\ket{\nu}\!\bra{\nu})_i$. It conserves all three populations and generalizes one-axis twisting to three modes with two independent twisting terms. For $r=1$, $\hat H_{\mathrm{int}}=J\hat N_2(N-\hat N_2)$ is OAT between $\ket2$ and the remaining population, whereas for $r\neq1$ the populations of $\ket1$ and $\ket3$ twist at different rates relative to $\ket2$, and the dynamics cannot be reduced to a single SU(2) OAT.

\begin{figure}[t]
    \centering
    \includegraphics[width=1\columnwidth]{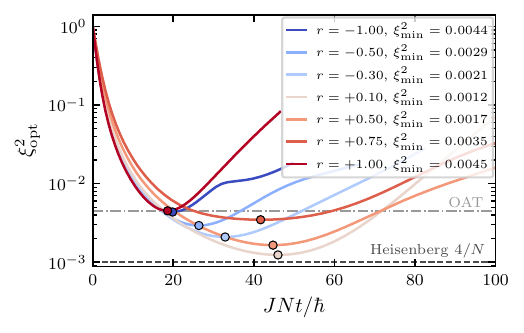}   
    \caption{Squeezing dynamics for all-to-all interactions ($\alpha=0$) at $N=4000$, from exact evolution in the permutationally symmetric subspace. Optimal squeezing parameter $\xi^2_{\mathrm{opt}}(t)$, optimized over linear $\mathrm{SU}(3)$ generators and readout operators, for the optimal initial product state $\big(\sqrt{p_1}\ket{1}+\sqrt{p_2}\ket{2}+\sqrt{p_3}\ket{3}\big)^{\otimes N}$ at each coupling ratio $r=J_2/J_1$. Legend gives $\xi^2_{\min}=\min_t\xi^2_{\mathrm{opt}}(t)$ and circles mark the minima. OAT $N^{-2/3}$ (dash-dotted line) and Heisenberg limit $4/N$ (dashed line). }
    \label{fig:ED_xi_vs_t}
\end{figure}
We now validate this analytical argument using exact numerics establishing OAT-type squeezing at $r=1$, show that squeezing is generated for generic coupling ratios $r$, and demonstrate the advantage of the three-level system ($r\neq 1$) over effective 2-level OAT squeezing.

In Fig.~\ref{fig:ED_xi_vs_t} we show the time evolution of the optimal squeezing parameter for a range of coupling ratios $r$ for all-to-all interactions and $N=4000$ spins. Time evolution is computed for the state that achieves the best squeezing at each $r$ (which is also $N$ dependent, see Appendix~\ref{app:readout}). We observe that the best squeezing strongly depends on the coupling ratio $r$, and is maximized at a value $r \neq 1$. Since dynamics at $r=1$ reduces to one-axis twisting (OAT) in an effective 2-level basis, this demonstrates a direct advantage of the 3-level system over effective $\mathrm{SU}(2)$ OAT squeezing, reaching a factor $3.64$ (5.6\,dB) at $N=4000$. Moreover, Fig.~\ref{fig:ED_xi_vs_t} illustrates a trade-off between the maximal metrological gain and the optimal squeezing time. While $r=0.1$ achieves substantially deeper squeezing, the minimum is reached at $JNt_{\mathrm{opt}}/\hbar\approx46$, compared with $18.6$ for OAT at $r=1$.

\begin{figure}[t]
    \centering
    \includegraphics[width=1\columnwidth]{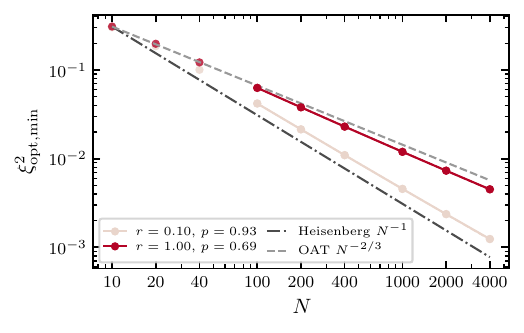}   
   \caption{Scaling of the minimal squeezing parameter $\xi^2_{\mathrm{opt,min}}=\min_t\xi^2_{\mathrm{opt}}(t)$ with system size $N$ for all-to-all interactions ($\alpha=0$) from exact calculations, at $r=1$ (OAT) and $r=0.1$, optimized over the initial product state at each $N$ (Table~\ref{tab:optimal_2}). Fits of $\xi^2_{\min}=AN^{-p}(1+bN^{-1/3})$ over $100\le N\le4000$ give $p=0.689(4)$ for $r=1$, identical to the same fit applied to the exact OAT result, and $p=0.934(7)$ for $r=0.1$ (fit-window dependence in Appendix~\ref{app:scaling}). Gray lines: guides to the eye $\propto N^{-2/3}$ (dashed) and $\propto N^{-1}$ (dash-dotted) through the $N=10$ data point.}\label{fig:ED_xi_vs_N}
\end{figure}

In Fig.~\ref{fig:ED_xi_vs_N} we study the scaling of the achievable squeezing with system size for the symmetric point $r=1$, which realizes OAT, and the asymmetric point $r=0.1$, which gives the strongest squeezing at large $N$. Over $100\le N\le4000$ we obtain $p=0.689(4)$ for $r=1$, which coincides with the exact OAT result fitted with the same ansatz and window (the asymptotic value $2/3$ is approached slowly), and a significantly larger exponent $p=0.934(7)$ for $r=0.1$. This demonstrates a scalable advantage of the three-level dynamics over OAT, at least over the range of system sizes accessible here, and at every system size an appropriately chosen $r\neq1$ outperforms OAT (Table~\ref{tab:optimal}).

Finally, we turn to the case of dipolar interactions to demonstrate that the same phenomenology persists there, establishing the scalable generation of spin squeezing in 3-level Rydberg arrays, as well as the advantage of 3-level over effective 2-level OAT dynamics.

\begin{figure}[t]
    \centering
    \includegraphics[width=1\columnwidth]{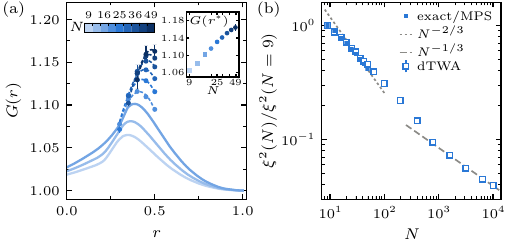}   
\caption{%
Metrological advantage of the optimal spin-1 product state (dipolar interactions, $L_x \times L_y$ lattice, open boundaries). 
(a)~$G(r)=\xi^2_{\mathrm{bright}}/\xi^2_{\mathrm{opt}}(r)$ of the optimal state at fixed $r$ relative to the bright state at $r=1$, for $N=9$--$49$ (color). Solid lines: ED ($N=9$, 12, 16). Symbols: MPS ($N=20$--$49$; two-site TDVP, $\chi\le512$, extrapolated to $\chi\to\infty$), with error bars combining the extrapolation and method errors (Appendix~\ref{app:dipolar_optimal_state}); dashed: quadratic fits. Inset: maximal gain $G(r^*)$ versus $N$ (squares: ED; circles: MPS). 
(b)~$\xi^2_{\min}$ at $r^*(N)$, normalized to its value at $N=9$: ED/MPS (filled squares) and dTWA ($10^4$ trajectories; open squares, errors smaller than symbol size), with $r^*$ and the initial state extrapolated for $N\ge49$ (Table~\ref{tab:sm_dtwa_states}). Gray lines: $N^{-2/3}$ through the $N = 49$ MPS point (dotted) and $N^{-1/3}$ (dashed, fitted to dTWA points for $N \ge 784$).}   \label{fig:dipolar_results}
\end{figure}

In Fig.~\ref{fig:dipolar_results}(a) we show the gain of the optimal three-level protocol over the effective spin-1/2 dynamics, $G(r)=\xi^2_{\mathrm{bright}}/\xi^2_{\mathrm{opt}}(r)$, where $\xi^2_{\mathrm{opt}}(r)$ is minimized over the initial product state and time at fixed $r$, and $\xi^2_{\mathrm{bright}}$ is the squeezing of the bright state at $r=1$, which equals that of the spin-1/2 dipolar XY model. %
For $N=9$ ($3\times3$), 12 ($3\times4$) and 16 ($4\times4$) we use exact time evolution (solid lines), and for larger systems up to $7\times7$ ($N=49$) MPS time evolution with the two-site time-dependent variational principle (TDVP)~\cite{Haegeman2011,Haegeman2016} in \textsc{ITensors.jl}~\cite{Itensor1,Itensor2}, extrapolated in the bond dimension $\chi\le512$ (Appendix~\ref{app:dipolar_optimal_state}). %
$G(r)$ exhibits a clear maximum at an intermediate ratio $r^*$, which increases from $r^*\approx0.35$ at $N=9$ to $r^*\approx0.48$ at $N=49$. The maximal gain (inset) increases from $0.27\,$dB (6.5\%) at $N=9$ to $0.66(3)\,$dB (16.5\%) at $N=49$. Thus the three-level advantage persists, and grows, over the accessible range of system sizes, although its growth slows with $N$.

In Fig.~\ref{fig:dipolar_results}(b) we show the scaling of the optimal squeezing at $r^*(N)$ from ED and MPS for $N\le49$ (filled symbols), and from dTWA up to $N=10^4$ (open symbols), for which we extrapolate $r^*$ and the optimal state to larger $N$ (Appendix~\ref{app:dipolar_optimal_state}). The squeezing increases steadily with $N$, reaching $\xi^2_{\min}\approx-18\,$dB at $N=10^4$ in dTWA, i.e.\ $\xi^2(N)/\xi^2(9)\approx0.04$ in Fig.~\ref{fig:dipolar_results}(b). %
 While the small system size results appear consistent with one-axis-twisting like scaling $N^{-2/3}$, at larger system sizes the scaling appears closer to $N^{-1/3}$. In any case, since the spin-1/2 XY state is contained in the optimized family, the optimal three-level squeezing is at least as good as that of the spin-1/2 dipolar XY model at every $N$, which is expected to be scalable in two dimensions~\cite{Comparin2022a,Roscilde2023,Block2024}.

Our results confirm that scalable squeezing persists in 2D dipolar arrays, and that the three-level protocol at $r\approx r^*$ outperforms the effective spin-1/2 dynamics of the bright state at $r=1$, which for dipolar interactions realizes the spin-1/2 XY model rather than OAT. The advantage of the full $\mathrm{SU}(3)$ manifold thus persists in realistic planar Rydberg tweezer geometries.

{\it Discussion and outlook.---}%
We have studied the dynamical generation of spin squeezing in arrays of dipole-interacting spin-1 (three-level) systems, realizable with Rydberg atoms in optical tweezer arrays, quantified by a generalized squeezing parameter optimized over all linear $\mathrm{SU}(3)$ generators and readout observables. %

For symmetric exchange couplings ($r=1$), the interaction conserves the numbers of bright and dark excitations, and the dynamics reduces to one-axis twisting of an effective two-level system. Away from this point the bright and dark manifolds are coupled, and the dynamics exploits the full three-level structure. For all-to-all interactions, the optimal protocol at $r\approx0.1$ improves the squeezing over OAT by a factor of 3.6 at $N=4000$, with a fitted exponent $p\approx0.93$ compared with $0.69$ for OAT over the same range of $N$, at the cost of a longer squeezing time. 
For 2D dipolar arrays, exact and tensor-network simulations up to $N=49$ show that the optimal three-level protocol, at $r^*\approx0.35$-$0.48$, outperforms the effective spin-1/2 dynamics by up to $0.66\,$dB, with a gain that grows with $N$ over the accessible range, and semiclassical simulations show that the squeezing keeps scaling up to the largest simulated size at $N=10^4$.

%
%
Our results represent a first step toward harnessing the richer internal structure of multi-level Rydberg systems for entanglement-enhanced metrology. The three-level model studied here is the simplest realization beyond a qubit and can naturally be implemented in current Rydberg tweezer experiments using the level schemes provided. %

A natural extension of this work is the study of higher-dimensional internal spaces ($D>3$), which may reveal further squeezing channels associated with higher-rank multipole operators. Additionally, the interplay between the distinct SU(2) subgroups identified here may enable multi-parameter estimation protocols, where simultaneous squeezing in different quadratures provides sensitivity to independent physical quantities. %
Finally, it will be interesting to explore whether Floquet engineering via optimized pulse sequences~\cite{Lukin_2020_Robust,PhysRevLett.131.220803,PhysRevA.108.053318,Scholl2022,Geier2021} extended to the multi-level setting can further enhance the achievable squeezing. While we have established here optimal product states, generators and read-out operations for the considered class of spin-exchange Hamiltonians, more complex interaction Hamiltonians may result in distinct squeezing mechanisms.

 \begin{acknowledgments}
This material is based upon work supported by the Air Force Office of Scientific Research under Award No.\ FA9550-25-1-0340. TB acknowledges support from the National Science Foundation through NRT NSF Grant No.\ DGE-2510202, as well as from the Department of Energy under DE-SC0026761. Some of the computing for this project was performed at the OU Supercomputing Center for Education \& Research (OSCER) at the University of Oklahoma (OU).  Some of the calculations were performed in the PETE system of the High Performance Computing Center at Oklahoma State University, NSF Grant No. OAC-1531128.

\end{acknowledgments}

\noindent\textit{Data availability}---The data that support the findings of this article are publicly available at \cite{data}.


\bibliography{main}{}

\cleardoublepage

\appendix


\counterwithout{equation}{section} \setcounter{equation}{0}
\setcounter{figure}{0}
\setcounter{table}{0}
\renewcommand{\theequation}{S\arabic{equation}}
\renewcommand{\thefigure}{S\arabic{figure}}
\renewcommand{\thetable}{S\arabic{table}}
\renewcommand{\theHfigure}{S\arabic{figure}}
\renewcommand{\theHtable}{S\arabic{table}}

\section*{Appendices}
The appendices contain the explicit Rydberg level schemes and an analysis of competing processes (Appendix~\ref{app:candidates}), the operator conventions (Appendix~\ref{app:operators}), the derivation of the bright/dark decomposition (Appendix~\ref{app:bright_dark}), the definition of the optimal generalized squeezing parameter (Appendix~\ref{app:xiopt}), extended all-to-all data and optimal protocols (Appendices~\ref{app:scaling} and \ref{app:readout}), details of the dipolar ED/MPS/dTWA simulations (Appendix~\ref{app:dipolar_optimal_state}), and benchmarks of dTWA against exact and MPS results (Appendix~\ref{app:benchmarks}).

\section{Candidate Rydberg configurations}
\label{app:candidates}
We provide full level schemes to realize the tunable Hamiltonian of the main
text with Rydberg levels, taking cesium as an example species. The three
qudit levels $|1\rangle,|2\rangle,|3\rangle$ are encoded in three
$(n,\ell,j,m_j)$ Rydberg states; resonant dipolar exchange between neighboring
atoms corresponds to the exchange coupling $J_1$ (on the $|1\rangle\!\leftrightarrow\!|2\rangle$
transition) and $J_2$ (on $|2\rangle\!\leftrightarrow\!|3\rangle$). Atomic
energies, dipole matrix elements and lifetimes are obtained from the
\textsc{arc} library~\cite{ARC_library}; lifetimes are evaluated at $T=300\,$K and include blackbody-induced transitions.

\begin{figure}[!t]
    \centering  
    \includegraphics[width=\linewidth]{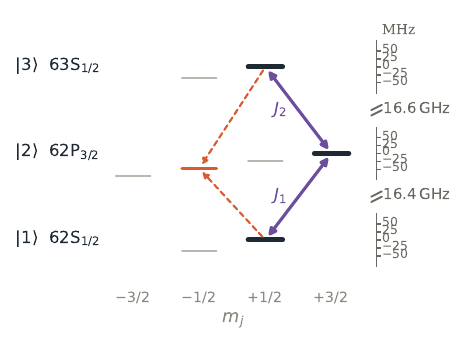}
    \caption{Illustration of level scheme (Cs), desired processes, and undesired processes. $|1\rangle \equiv |62\mathrm{S}_{1/2}^{+1/2}\rangle$, $|2\rangle \equiv |62\mathrm{P}_{3/2}^{+3/2}\rangle$, $|3\rangle \equiv |63\mathrm{S}_{1/2}^{+1/2}\rangle$, and leakage state $|\tilde{2}\rangle \equiv |62\mathrm{P}_{3/2}^{-1/2}\rangle$, at $B = 12.6\,\mathrm{G}$ and $R = 12\,\mu\mathrm{m}$. %
    The qutrit levels are indicated as thick black, other levels in the same manifold as gray, desired exchange in purple, undesired exchange partners in red. Energy axis linear in MHz relative to the qutrit level in each manifold, and broken between manifolds with the gap indicated on the axis. 
Desired exchange. $|3,2\rangle \leftrightarrow |2,3\rangle$ ($J_2/2\pi = -1.31\,\mathrm{MHz}$); $|2,1\rangle \leftrightarrow |1,2\rangle$ ($J_1/2\pi = -1.37\,\mathrm{MHz}$). 
Undesired processes. $|3,2\rangle \rightarrow |\tilde{2},3\rangle$; 
$|2,1\rangle \rightarrow |1,\tilde{2}\rangle$; 
$|2,2\rangle \rightarrow |3,1\rangle$. 
    \label{fig:fig1Threelevelatomicarray_wide}}
\end{figure}

We illustrate one of the resulting schemes in Fig.~\ref{fig:fig1Threelevelatomicarray_wide}, corresponding to the near-symmetric configuration $r = J_2/J_1 = 0.954$ encoded in the cesium states $\ket{1} = \ket{62S_{1/2}, m_j=+1/2}$, $\ket{2} = \ket{62P_{3/2}, m_j=+3/2}$, and $\ket{3} = \ket{63S_{1/2}, m_j=+1/2}$ at interatomic spacing $R = 12\,\mu\mathrm{m}$ and magnetic field $B = 12.6\,\mathrm{G}$, which Zeeman-detunes competing exchange channels to $|\delta| = 47.1\,\mathrm{MHz} \gg |J_{1,2}|/2\pi$.

\subsection{Selection procedure}
To find suitable candidates we first enumerate consecutive $(n,\ell,j,m_j)$ level schemes with $|n_\nu-n_{\nu+1}|\le1$ and evaluate the spacing-independent quantities: the coupling ratio $J_2/J_1$~\eqref{eq:ratio}, the $\pi/\sigma$
character of the transitions, the lifetimes of the states, and the F\"orster defect $\Delta_F=|E_1+E_3-2E_2|$. We then check for leakage out of the qutrit manifold via time-evolution of the full two-body Hamiltonian to find optimal level schemes with suppressed leakage (Appendix~\ref{app:neglected}).

 Table~\ref{table:param} lists every level with its lifetime and each transition with its coupling constant $C_3/h$ and resulting $J_i/2\pi$. All presented schemes use low-$\ell$ $S$, $P$ and
$D$ states.  Table~\ref{tab:audit} provides quantitative estimates of additional processes (see Appendix~\ref{app:audit}) for the same level schemes.

\subsection{Dipolar exchange \label{app:dipolar_exchange}}
The atoms occupy a two-dimensional array in the $x$--$y$ plane with the
quantization field $\mathbf B\parallel\hat{\mathbf z}$, so every bond makes an
angle $\theta=\pi/2$ with $\mathbf B$. The dipole--dipole interaction is
\begin{equation}
  \hat H_{\rm dd}=\frac{1}{4\pi\epsilon_0 R^3}
  \Big[\hat{\mathbf d}_A\!\cdot\!\hat{\mathbf d}_B
       -3(\hat{\mathbf d}_A\!\cdot\!\hat{\mathbf n})
         (\hat{\mathbf d}_B\!\cdot\!\hat{\mathbf n})\Big],
  \label{eq:vdd}
\end{equation}
with $R$ the interatomic separation and $\hat{\mathbf n}$ the bond direction.

Projecting Eq.~\eqref{eq:vdd} onto the qudit levels, the
resonant exchange interaction strength is
\begin{equation}
  \frac{J_i}{2\pi}=\kappa\, \mathcal{A}_i\,\frac{|M_i|^2}{R^3},
  \label{eq:Ji}
\end{equation}
with  $\kappa=\frac{e^2a_0^2}{4\pi\epsilon_0 h}$, $M_i=\langle i|\hat d_q|i+1\rangle/(ea_0)$ the dimensionless transition
dipole matrix element, and the angular coefficient $\mathcal{A}_i=+1$ for a $\pi$ transition ($\Delta m_j=0$) and
$\mathcal{A}_i=-\tfrac12$ for a $\sigma$ transition ($|\Delta m_j|=1$). 

In terms of $C_3^{(i)}/h= \kappa \, \mathcal{A}_i \,|M_i|^2$ we have $J_i/(2\pi)=(C_3^{(i)}/h)/R^3$. Because both couplings share the factor $R^{-3}$
and the angular factor $(1-3\cos^2\theta)$, the ratio
\begin{equation}
  \frac{J_2}{J_1}=\frac{\mathcal{A}_2|M_2|^2}{\mathcal{A}_1|M_1|^2}
  \label{eq:ratio}
\end{equation}
is fixed by the atomic structure and chosen levels. Its \emph{sign} is set by the $\pi/\sigma$ character of the two transitions (chosen
through $m_j$), and its \emph{magnitude} by $|\mathcal{A}_2/\mathcal{A}_1|\,|M_2|^2/|M_1|^2$. 
Together these give access to the broad range of $J_2/J_1$ tabulated below.

\subsection{Neglected interactions and leakage}
\label{app:neglected}
The leakage and level-mixing quantities are obtained by exact diagonalization of
the two-atom Hamiltonian in the full magnetic-sublevel basis of the three qudit
manifolds and additional levels dipole-coupled to the qutrit levels.

We characterize leakage out of the qutrit manifold by the worst case leakage
\begin{equation}
  \mathcal{L}=1-\min_{t\le T}\langle\psi(t)|\hat P|\psi(t)\rangle,
  \label{eq:leak}
\end{equation}
where $\hat{P}$ is the projector onto the 9-dimensional two-atom qutrit states, $T=\min\!\Big(\tfrac{15\hbar}{\max_i|J_i|},\tau_{\min}\Big)$, and we consider the maximum over initial two-atom states $|\psi(0)\rangle\in\{|1,2\rangle,|2,2\rangle,|2,3\rangle\}$. We report $B_{\rm req}$ as the smallest field at which $\max\mathcal{L}<1\%$.

\subsection{Additional processes \label{app:audit}}

Table~\ref{tab:audit} quantifies for each candidate the required B-fields and additional processes: 

\begin{itemize}
    \item[(a)] \textbf{Van der Waals shifts}: Second-order $C_6/R^6$ shifts of the $|2,2\rangle$ state reach at most $0.16$~MHz, i.e. $\lesssim 11.7\%$ of $J$.
    \item[(b)] \textbf{Off-resonant F\"orster channels}: Defect energies $\Delta_F = 0.22\text{--}34.2\,\mathrm{GHz}$ limit admixed pair population to $P_{\mathrm F}=(V_F/\Delta_F)^2\le3.3\times10^{-4}$, where $V_F$ is the $\ket{2,2}\to\ket{1,3}$ matrix element of Eq.~\eqref{eq:vdd} at $\theta=\pi/2$ (it vanishes for schemes combining a $\pi$ and a $\sigma$ transition, since couplings with $\Delta M=\pm1$ vanish at $\theta=\pi/2$). Summed over $\ket{1,3}$ and $\ket{3,1}$, the admixture is at most $6.7\times10^{-4}$, and the Zeeman shifts at $B_{\rm req}$ further increase $\Delta_F$, so these zero-field values are conservative.

    \item[(c)] \textbf{Additional exchange/leakage}: We quantify leakage out of the selected qutrit manifold by explicit evolution of the two-atom Hamiltonian in an extended set of states (see Appendix~\ref{app:neglected}). Leakage into other states is suppressed to below $1\%$ by moderate applied B-fields ($B_{\mathrm{req}}$). 
    \item[(d)] \textbf{Differential light shifts}: For a standard $1064\,\mathrm{nm}$ tweezer ($1\,\mu\mathrm{m}$ waist, $10\,\mathrm{mW}$), differential Stark shifts $\delta_{\mathrm{LS}} \approx 1.3\text{--}2.5\,\mathrm{MHz}$ cancel exactly in pair-preserving flip-flop processes on a homogeneous lattice. Assuming a $1\%$ intensity variation ($\delta I/I = 0.01$), the residual on-site disorder is $\delta_{\mathrm{LS}}(\delta I/I) \le 0.03\,\mathrm{MHz}$ ($\lesssim 3.3\%$ of $J$), which vanishes completely if tweezer light is off during interaction dynamics.
    \item[(e)] \textbf{Decay and coherence}: The number of exchange cycles within the shortest lifetime, $Q=(|J|/2\pi)\,\tau_{\min}$, ranges from 59 to 232, i.e.\ $|J|\tau_{\min}/\hbar\approx370$--$1460$. The optimal squeezing times of the dipolar arrays for $N\le49$ are $Jt_{\mathrm{opt}}/\hbar\approx1.05$--$1.15$ for the optimal states at $r^*$ ($0.41$--$0.64$ for the bright reference) and $\le1.5$ for all simulated $r$ (Appendix~\ref{app:dipolar_optimal_state}), so that the probability for an individual atom to decay before $t_{\mathrm{opt}}$ is $\lesssim4\times10^{-3}$ ($\approx10^{-3}$ for the $J_2/J_1=0.448$ scheme).
    \item[(f)] \textbf{Magnetic field mixing}: At $B_{\mathrm{req}}$, Paschen--Back admixture of the qudit states is $\le 2.4\times10^{-2}$ and shifts $J_2/J_1$ by $\delta r \le 0.018$ from its zero-field value.
    \item[(g)] \textbf{Positional disorder}: Defect-sorted tweezer arrays operate at
    $\gtrsim98$--$99\%$ filling and with residual positional disorder $\delta R/R\sim1$--$2\%$, giving
    weak bond disorder $\delta J/J=3\,\delta R/R\sim3$--$6\%$. This level of disorder is well below expected thresholds to suppress squeezing in disordered power-law interacting systems \cite{Kaplan-Lipkin2025,Begg2026,Solfanelli2026}.
\end{itemize}

\begin{table*}[tb!]
\newcommand{\midrow}[1]{\raisebox{-\normalbaselineskip}[0pt][0pt]{#1}}
\footnotesize
\begin{ruledtabular}
\begin{tabular}{ccccccc}
Configuration & State & Lifetime~$\left(\rm \mu s \right)$ & Level   & $|C_{3}|$/h$~\left(\rm MHz \cdot \mu m^3 \right)$ & Spacing~$\left(\rm \mu m \right)$ &  $J_i/2\pi~(\rm MHz)$  \\
\hline
& $|3\rangle$ & 153 & 60P$_{3/2,\, m_j=\tfrac{3}{2}}$ & \midrow{ 2064 } & &  \midrow{  $J_2/{2\pi}$ = -1.194 } \\
$J_2/J_1 =-0.844$ & $|2\rangle$ & 95 & 60S$_{1/2,\, m_j=\tfrac{1}{2}}$ & \midrow{ 2446 } & 12 & \midrow{$ J_1/{2\pi}$ = 1.415 } \\
& $|1\rangle$ & 147 & 59P$_{3/2,\, m_j=\tfrac{1}{2}}$ & &     &   \\
\hline
& $|3\rangle$ & 169 & 62P$_{1/2,\, m_j=\tfrac{1}{2}}$ & \midrow{ 1378 } & &  \midrow{  $J_2/{2\pi}$ = 0.5022 } \\
$J_2/J_1 =-0.609$ & $|2\rangle$ & 107 & 63S$_{1/2,\, m_j=\tfrac{1}{2}}$ & \midrow{ 2265 } & 14 & \midrow{$ J_1/{2\pi}$ = -0.8253 } \\
& $|1\rangle$ & 165 & 62P$_{3/2,\, m_j=\tfrac{3}{2}}$ & &     &   \\
\hline
& $|3\rangle$ & 169 & 62P$_{1/2,\, m_j=-\tfrac{1}{2}}$ & \midrow{ 1378 } & &  \midrow{  $J_2/{2\pi}$ = -0.5022 } \\
$J_2/J_1 =-0.456$ & $|2\rangle$ & 107 & 63S$_{1/2,\, m_j=\tfrac{1}{2}}$ & \midrow{ 3019 } & 14 & \midrow{$ J_1/{2\pi}$ = 1.1 } \\
& $|1\rangle$ & 165 & 62P$_{3/2,\, m_j=\tfrac{1}{2}}$ & &     &   \\
\hline
& $|3\rangle$ & 84 & 62D$_{3/2,\, m_j=-\tfrac{3}{2}}$ & \midrow{ 1085 } & &  \midrow{  $J_2/{2\pi}$ = 0.3953 } \\
$J_2/J_1 =-0.201$ & $|2\rangle$ & 171 & 63P$_{3/2,\, m_j=-\tfrac{3}{2}}$ & \midrow{ 5408 } & 14 & \midrow{$ J_1/{2\pi}$ = -1.971 } \\
& $|1\rangle$ & 85 & 62D$_{5/2,\, m_j=-\tfrac{5}{2}}$ & &     &   \\
\hline
& $|3\rangle$ & 84 & 62D$_{3/2,\, m_j=-\tfrac{1}{2}}$ & \midrow{ 361.6 } & &  \midrow{  $J_2/{2\pi}$ = -0.1318 } \\
$J_2/J_1 =0.149$ & $|2\rangle$ & 171 & 63P$_{3/2,\, m_j=-\tfrac{3}{2}}$ & \midrow{ 2424 } & 14 & \midrow{$ J_1/{2\pi}$ = -0.8832 } \\
& $|1\rangle$ & 112 & 64S$_{1/2,\, m_j=-\tfrac{1}{2}}$ & &     &   \\
\hline
& $|3\rangle$ & 103 & 62S$_{1/2,\, m_j=\tfrac{1}{2}}$ & \midrow{ 1286 } & &  \midrow{  $J_2/{2\pi}$ = 0.7442 } \\
$J_2/J_1 =0.248$ & $|2\rangle$ & 163 & 61P$_{1/2,\, m_j=\tfrac{1}{2}}$ & \midrow{ 5181 } & 12 & \midrow{$ J_1/{2\pi}$ = 2.998 } \\
& $|1\rangle$ & 77 & 60D$_{3/2,\, m_j=\tfrac{1}{2}}$ & &     &   \\
\hline
& $|3\rangle$ & 99 & 61S$_{1/2,\, m_j=-\tfrac{1}{2}}$ & \midrow{ 1970 } & &  \midrow{  $J_2/{2\pi}$ = -1.14 } \\
$J_2/J_1 =0.448$ & $|2\rangle$ & 153 & 60P$_{3/2,\, m_j=-\tfrac{3}{2}}$ & \midrow{ 4397 } & 12 & \midrow{$ J_1/{2\pi}$ = -2.545 } \\
& $|1\rangle$ & 74 & 59D$_{5/2,\, m_j=-\tfrac{5}{2}}$ & &     &   \\
\hline
& $|3\rangle$ & 169 & 62P$_{1/2,\, m_j=-\tfrac{1}{2}}$ & \midrow{ 1378 } & &  \midrow{  $J_2/{2\pi}$ = -0.3364 } \\
$J_2/J_1 =0.609$ & $|2\rangle$ & 107 & 63S$_{1/2,\, m_j=\tfrac{1}{2}}$ & \midrow{ 2265 } & 16 & \midrow{$ J_1/{2\pi}$ = -0.5529 } \\
& $|1\rangle$ & 165 & 62P$_{3/2,\, m_j=\tfrac{3}{2}}$ & &     &   \\
\hline
& $|3\rangle$ & 157 & 60P$_{1/2,\, m_j=\tfrac{1}{2}}$ & \midrow{ 1485 } & &  \midrow{  $J_2/{2\pi}$ = -0.541 } \\
$J_2/J_1 =0.809$ & $|2\rangle$ & 95 & 60S$_{1/2,\, m_j=-\tfrac{1}{2}}$ & \midrow{ 1834 } & 14 & \midrow{$ J_1/{2\pi}$ = -0.6685 } \\
& $|1\rangle$ & 147 & 59P$_{3/2,\, m_j=-\tfrac{3}{2}}$ & &     &   \\
\hline
& $|3\rangle$ & 107 & 63S$_{1/2,\, m_j=\tfrac{1}{2}}$ & \midrow{ 2265 } & &  \midrow{  $J_2/{2\pi}$ = -1.311 } \\
$J_2/J_1 =0.954$ & $|2\rangle$ & 165 & 62P$_{3/2,\, m_j=\tfrac{3}{2}}$ & \midrow{ 2373 } & 12 & \midrow{$ J_1/{2\pi}$ = -1.373 } \\
& $|1\rangle$ & 103 & 62S$_{1/2,\, m_j=\tfrac{1}{2}}$ & &     &   \\
\end{tabular}
\end{ruledtabular}
\caption{\label{table:param}
Summary of candidate Rydberg states of Cs to achieve target values of $J_2/J_1$ computed using the ARC library \cite{ARC_library}. We provide the lifetimes (at $300\,$K, including blackbody radiation), the specific levels, the magnitudes $|C_3|/h$ for the 3-2 and 2-1 transitions (the sign of $J_i$ is that of $\mathcal{A}_i$), and the resulting exchange couplings $J_i$ at the specified spacing.
}
\end{table*}

\begin{table*}[tb!]
\footnotesize
\setlength{\tabcolsep}{4pt}
\begin{ruledtabular}
\begin{tabular}{clcccccccc}
\multicolumn{3}{c}{Candidate} & (a) & (b) & (c) & (d) & (e) & \multicolumn{2}{c}{(f)} \\
$J_2/J_1$ & States ($\ket{3}$,$\ket{2}$,$\ket{1}$) & $J/2\pi$~(MHz) & $\delta_{\rm vdW}/J$ & $P_{\rm F}$ & $B_{\rm req}$~(G) & $\delta_{\rm LS}$~(MHz) & $Q$ & $w_{\rm mix}$ & $\delta r$ \\
\hline
-0.844 & 60P$_{3/2}^{+3/2}$, 60S$_{1/2}^{+1/2}$, 59P$_{3/2}^{+1/2}$ & 1.42 & 0.025 & 0 & 14 & 1.51 & 134 & $5.2\times10^{-5}$ & 0.008 \\
-0.609 & 62P$_{1/2}^{+1/2}$, 63S$_{1/2}^{+1/2}$, 62P$_{3/2}^{+3/2}$ & 0.825 & 0.031 & 0 & 8 & 1.84 & 89 & $2.4\times10^{-5}$ & 0.009 \\
-0.456 & 62P$_{1/2}^{-1/2}$, 63S$_{1/2}^{+1/2}$, 62P$_{3/2}^{+1/2}$ & 1.1 & 0.023 & 0 & 11 & 1.84 & 118 & $4.7\times10^{-5}$ & 0.009 \\
-0.201 & 62D$_{3/2}^{-3/2}$, 63P$_{3/2}^{-3/2}$, 62D$_{5/2}^{-5/2}$ & 1.97 & 0.038 & 0 & 11 & 1.35 & 167 & $4.8\times10^{-4}$ & 0.018 \\
0.149 & 62D$_{3/2}^{-1/2}$, 63P$_{3/2}^{-3/2}$, 64S$_{1/2}^{-1/2}$ & 0.883 & 0.085 & $3.1\times10^{-9}$ & 8 & 2.12 & 75 & $3.9\times10^{-4}$ & 0.007 \\
0.248 & 62S$_{1/2}^{+1/2}$, 61P$_{1/2}^{+1/2}$, 60D$_{3/2}^{+1/2}$ & 3 & 0.0039 & $4.4\times10^{-9}$ & 82 & 2.52 & 232 & $2.4\times10^{-2}$ & 0.006 \\
0.448 & 61S$_{1/2}^{-1/2}$, 60P$_{3/2}^{-3/2}$, 59D$_{5/2}^{-5/2}$ & 2.54 & 0.046 & $6.1\times10^{-9}$ & 14 & 1.52 & 189 & 0 & 0 \\
0.609 & 62P$_{1/2}^{-1/2}$, 63S$_{1/2}^{+1/2}$, 62P$_{3/2}^{+3/2}$ & 0.553 & 0.021 & $1.6\times10^{-10}$ & 5 & 1.84 & 59 & $9.9\times10^{-6}$ & 0.003 \\
0.809 & 60P$_{1/2}^{+1/2}$, 60S$_{1/2}^{-1/2}$, 59P$_{3/2}^{-3/2}$ & 0.668 & 0.021 & $6.1\times10^{-8}$ & 6 & 1.88 & 63 & $1.1\times10^{-5}$ & 0.004 \\
0.954 & 63S$_{1/2}^{+1/2}$, 62P$_{3/2}^{+3/2}$, 62S$_{1/2}^{+1/2}$ & 1.37 & 0.12 & $3.3\times10^{-4}$ & 13 & 1.48 & 141 & 0 & 0 \\
\end{tabular}
\end{ruledtabular}
\caption{\label{tab:audit} Quantitative characterization of additional effects for each candidate level configuration. $J=\max(|J_1|,|J_2|)$ sets the exchange scale. (a) $\delta_{\rm vdW}/J$: second-order van der Waals shift of $|2,2\rangle$ over $J$; (b) $P_{\rm F}$: off-resonant F\"orster population admixed into $|1,3\rangle$, $(V_F/\Delta_F)^2$ (zero for mixed $\pi$/$\sigma$ schemes); (c) $B_{\rm req}$: field for $<\!1\%$ worst-channel leakage; (d) $\delta_{\rm LS}$: max differential light shift between sites assuming a $\lambda=1064$~nm, $1\,\mu$m-waist, $10$~mW tweezer; it cancels exactly in the level-preserving exchange and enters only through intensity disorder, giving an on-site detuning $\delta_{\rm LS}\,(\delta I/I)\lesssim 0.03$~MHz at $\delta I/I=1\%$;  (e) $Q=(|J|/2\pi)\,\tau_{\min}$: exchange cycles within the shortest lifetime; (f) $w_{\rm mix}$ / $\delta r$: Paschen--Back admixture of the qudit states at $B_{\rm req}$ and the induced drift of $r=J_2/J_1$ from its zero-field value.}
\end{table*}

\cleardoublepage

\section{Operator Conventions \label{app:operators}}
In this section we provide the explicit conventions used in this work for the spin-quadrupolar as well as Gell-Mann operators.

\subsection{Definitions of the spin-quadrupolar operators}
The three levels are identified with the magnetic sublevels of a spin-1 degree of freedom as $\ket{1}\equiv\ket{m_s{=}{+}1}$, $\ket{2}\equiv\ket{m_s{=}0}$, $\ket{3}\equiv\ket{m_s{=}{-}1}$, in which basis the single-particle spin operators read
\begin{align}
\hat s_x&=\frac{1}{\sqrt2}\begin{pmatrix}0&1&0\\1&0&1\\0&1&0\end{pmatrix},\; \quad
\hat s_y=\frac{1}{\sqrt2}\begin{pmatrix}0&-i&0\\i&0&-i\\0&i&0\end{pmatrix},\; \\
\hat s_z & =\begin{pmatrix}1&0&0\\0&0&0\\0&0&-1\end{pmatrix}.
\end{align}
The quadrupole (nematic) operators are defined as~\cite{Yukawa2013}
\begin{equation}
\begin{aligned}
\hat q_{ab}&=\hat s_a\hat s_b+\hat s_b\hat s_a \quad (a\neq b),\\
\hat d_{xy}&=\hat s_x^2-\hat s_y^2 = \ket{1}\!\bra{3}+\ket{3}\!\bra{1},\\
\hat y&=\frac{1}{\sqrt3}\left(2\hat{s}_z^2-\hat{s}_x^2 - \hat{s}_y^2 \right),
\end{aligned}
\label{SI-eq:quadrupoleDefs}
\end{equation}
where lower-case symbols denote single-particle operators. The corresponding collective operators are defined as sums over sites, e.g., $\hat S_x=\sum_i \hat s_{x,i}$, $\hat Q_{yz}=\sum_i\hat q_{yz,i}$, and are denoted by capital letters.

\subsection{Definition of Gell-Mann Matrices}
\label{sec:gell_mann}
In our treatment of spin-1 squeezing below, it is more convenient to expand all operators in terms of the SU(3) group generators. We utilize the standard basis of the eight Gell-Mann matrices, $\lambda_k$ ($k=1,\dots,8$). These are traceless, Hermitian $3 \times 3$ matrices that satisfy the orthogonality condition $\text{Tr}(\lambda_j \lambda_k) = 2 \delta_{jk}$. These are explicitly given by
\begin{align}
\hat \lambda_1 &= 
\begin{pmatrix} 
0 & 1 & 0 \\ 
1 & 0 & 0 \\ 
0 & 0 & 0 
\end{pmatrix}, &
\hat \lambda_2 &= 
\begin{pmatrix} 
0 & -i & 0 \\ 
i & 0 & 0 \\ 
0 & 0 & 0 
\end{pmatrix}, \nonumber \\
\hat \lambda_3 &= 
\begin{pmatrix} 
1 & 0 & 0 \\ 
0 & -1 & 0 \\ 
0 & 0 & 0 
\end{pmatrix}, &
\hat \lambda_4 &= 
\begin{pmatrix} 
0 & 0 & 1 \\ 
0 & 0 & 0 \\ 
1 & 0 & 0 
\end{pmatrix}, \nonumber \\
\hat \lambda_5 &= 
\begin{pmatrix} 
0 & 0 & -i \\ 
0 & 0 & 0 \\ 
i & 0 & 0 
\end{pmatrix}, &
\hat \lambda_6 &= 
\begin{pmatrix} 
0 & 0 & 0 \\ 
0 & 0 & 1 \\ 
0 & 1 & 0 
\end{pmatrix}, \nonumber \\
\hat \lambda_7 &= 
\begin{pmatrix} 
0 & 0 & 0 \\ 
0 & 0 & -i \\ 
0 & i & 0 
\end{pmatrix}, &
\hat \lambda_8 &= \frac{1}{\sqrt{3}}
\begin{pmatrix} 
1 & 0 & 0 \\ 
0 & 1 & 0 \\ 
0 & 0 & -2 
\end{pmatrix}.
\label{eq:SI_Gell-Mann}
\end{align}
and we again denote collective operators with capital symbols, i.e. $\hat{\Lambda}_a = \sum_i \hat{\lambda}_{a,i}$.

\section{3-level Model and Bright Manifold OAT}\label{app:bright_dark}

The three-level Hamiltonian is
\begin{equation}
    \begin{aligned}    
    \hat{H}_{\Omega} = &\sum_{i} \bigg[ \frac{\Omega_{1}}{2} \left( \hat{T}_{i}^{12} + \hat{T}_{i}^{21} \right)  + \frac{\Omega_{2}}{2}  \left( \hat{T}_{i}^{23} + \hat{T}_{i}^{32} \right) \bigg],\\
    \hat{H}_{\text{int}} = & \sum_{i \neq j} \bigg[ J_{ij}^{1} \hat{T}^{21}_{i} \hat{T}^{12}_{j} + J^{2}_{ij} \hat{T}^{32}_{i} \hat{T}^{23}_{j} \bigg],
    \end{aligned} 
    \label{SI-eq:ThreelevelFullHamiltonian}
\end{equation}
which we note is Hermitian since the sum over all pairs $i\neq j$ together with $J^\nu_{ij}=J^\nu_{ji}$ implies that the $(j,i)$ term is the Hermitian conjugate of the $(i,j)$ term. 

We consider a range of coupling ratios $r =J_{2}/J_{1}$, where $r=1$ corresponds to symmetric interactions $J_1 = J_2$, and due to the symmetry under $r \rightarrow 1/r$ by exchanging the state labels $1 \leftrightarrow 3$, considering $|r| \le 1$ is sufficient.

We may rewrite the interaction Hamiltonian in Eq.~\eqref{SI-eq:ThreelevelFullHamiltonian} using the basis composed of $\left\{\ket{b},\ket{d},\ket{2}\right\}$ with
\[\ket{b}=\frac{\ket{1} +\ket{3}}{\sqrt{2}}, \qquad \ket{d}=\frac{\ket{1}-\ket{3}}{\sqrt{2}},
\] and the manifold ladder operators $\hat b^\pm_i$, $\hat d^\pm_i$ defined by $\hat b^+_i=(\ket{b}\!\bra{2})_i$, $\hat d^+_i=(\ket{d}\!\bra{2})_i$ and $\hat b^-_i=(\hat b^+_i)^\dagger$, $\hat d^-_i=(\hat d^+_i)^\dagger$. Inverting the basis change, $\ket{1}=(\ket{b}+\ket{d})/\sqrt2$ and $\ket{3}=(\ket{b}-\ket{d})/\sqrt2$, the two exchange terms become
\begin{equation}
\begin{aligned}
\hat{T}^{21}_i\hat{T}^{12}_j&=\tfrac12\big(\hat b^-_i+\hat d^-_i\big)\big(\hat b^+_j+\hat d^+_j\big),\\
\hat{T}^{32}_i\hat{T}^{23}_j&=\tfrac12\big(\hat b^+_i-\hat d^+_i\big)\big(\hat b^-_j-\hat d^-_j\big).
\end{aligned}
\label{SI-eq:BDtransformation}
\end{equation}
Summing over pairs $i\neq j$ with $J_{ij}=J_{ji}$ and relabeling $i\leftrightarrow j$, we obtain
\begin{equation}
\begin{aligned}
    \hat{H}_{\text{int}}&= \sum_{i\neq j} \frac{J^1_{ij} + J^2_{ij}}{2}( \hat{b}^{+}_{i} \hat{b}^{-}_{j} +\hat{d}^{+}_{i}\hat{d}^{-}_{j}) \\
    &+  \sum_{i\neq j}\frac{J^1_{ij} - J^2_{ij}}{2}( \hat{b}^{+}_{i} \hat{d}^{-}_{j} +\hat{d}^{+}_{i}\hat{b}^{-}_{j})
    \label{SI-eq:InteractionHamiltonianw_general_BD}
\end{aligned}
\end{equation}
For $J^1_{ij}\neq J^2_{ij}$ the second line converts a bright excitation on one site into a dark excitation on another, so that $\hat N_B$ and $\hat N_D$ are no longer separately conserved (only $\hat N_1$, $\hat N_2$, $\hat N_3$ are), and the dynamics explores the full three-level space even for an initially bright state.

In the symmetric case it reduces to
\begin{equation}
\begin{aligned}
    \hat{H}_{\text{int}}&= \sum_{i\neq j} J_{ij}( \hat{b}^{+}_{i} \hat{b}^{-}_{j} +\hat{d}^{+}_{i}\hat{d}^{-}_{j})
    \label{SI-eq:InteractionHamiltonianwBD}
\end{aligned}
\end{equation}
This Hamiltonian acts within two SU(2) manifolds, the bright manifold with generators $\hat{\boldsymbol B}
=\{\tfrac12\hat{S}_{x},\tfrac12\hat{Q}_{yz},\tfrac14(\hat{D}_{xy}+ \sqrt{3}\hat{Y})\}$ and a dark manifold with generators $\hat{\boldsymbol D}=\{\tfrac12\hat{Q}_{xz},\tfrac12\hat{S}_{y},\tfrac14(\sqrt{3} \hat{Y}-\hat{D}_{xy})\}$. At the single-particle level these are precisely the Pauli-type operators of the two effective two-level systems,
\begin{equation}
\begin{aligned}
\hat b^x_i&=\tfrac12(\hat b^+_i+\hat b^-_i)=\tfrac12 \hat s_{x,i},\\
\hat b^y_i&=\tfrac{1}{2i}(\hat b^+_i-\hat b^-_i)=\tfrac12 \hat q_{yz,i},\\
\hat b^z_i&=\tfrac12\big(\ket{b}\!\bra{b}-\ket{2}\!\bra{2}\big)_i=\tfrac14\big(\hat d_{xy,i}+\sqrt3\,\hat y_i\big),
\end{aligned}
\label{SI-eq:brightGenerators}
\end{equation}
and analogously $\hat d^x_i=\tfrac12\hat q_{xz,i}$, $\hat d^y_i=\tfrac12\hat s_{y,i}$, $\hat d^z_i=\tfrac12(\ket{d}\!\bra{d}-\ket{2}\!\bra{2})_i=\tfrac14(\sqrt3\,\hat y_i-\hat d_{xy,i})$ for the dark manifold, with collective operators $\hat B^a=\sum_i\hat b^a_i$ and $\hat D^a=\sum_i\hat d^a_i$. These satisfy the $\mathfrak{su}(2)$ commutation relations $[\hat b^a_i,\hat b^b_i]=i\epsilon_{abc}\hat b^c_i$ and $[\hat d^a_i,\hat d^b_i]=i\epsilon_{abc}\hat d^c_i$, which follow directly from Eq.~\eqref{SI-eq:brightGenerators}, and each generator annihilates the third level ($\hat b^a_i\ket{d}_i=0$, $\hat d^a_i\ket{b}_i=0$).

The interaction has the  form of a one-axis twisting (OAT) Hamiltonian,
\begin{equation}
\hat{H}=\sum_{i\neq j} J_{ij}\, \hat{b}_{i}^{+}\hat{b}_{j}^{-} = J\big[\hat{\boldsymbol B}^2 - (\hat{B}^{z})^{2}\big] - \tfrac{J}{2}N ,
\end{equation}
in the limit of all-to-all interactions, $J_{ij}=J$, within the bright sector ($\hat N_D=0$). This follows from $\sum_{i\neq j}\hat b^+_i\hat b^-_j=\hat B^+\hat B^--\sum_i\hat b^+_i\hat b^-_i$ with $\hat B^+\hat B^-=\hat{\boldsymbol B}^2-(\hat B^z)^2+\hat B^z$ and $\sum_i\hat b^+_i\hat b^-_i=N/2+\hat B^z$. The linear terms cancel, and the restricted sum generates the same OAT dynamics as the unrestricted one.

Since the number of bright and dark excitations in Eq.~\eqref{SI-eq:InteractionHamiltonianwBD}, $\hat N_B=\sum_i\ket{b}\!\bra{b}_i$ and $\hat N_D=\sum_i\ket{d}\!\bra{d}_i$, are individually conserved, the dynamics in the two manifolds effectively decouple. In particular, for an initial product state with no dark (bright) component the dynamics remain confined to the bright (dark) two-level subspace at every site. Thus, the initial state $\ket{\psi}_{I}=\ket{b_{x}}^{\otimes N}$ with $\ket{b_x}=(\ket{b}+\ket{2})/\sqrt2$ leads to OAT type squeezing in the $\hat{\mathbf{B}}$ manifold, similarly $\ket{\psi}_{I}=\ket{d_x}^{\otimes N}$ with $\ket{d_x}=(\ket{d}+\ket{2})/\sqrt2$ would squeeze in the $\hat{\mathbf{D}}$ manifold.

\section{Optimal generalized Squeezing Parameter \label{app:xiopt}}

In this section we define the generalized squeezing parameter optimized over the complete basis of linear generator and read-out operators of SU(D), applied to spin 1, i.e. SU(3) \cite{bahamnia2026gapprotectedheisenberglimitedsqueezinglocally}.

We follow the methodology of Ref.~\cite{Gessner2019}, taking as our accessible set the collective SU(3) operators of $N$ spin-1 particles. These operators are normalized such that $\mathrm{tr}(\hat\lambda_a\hat\lambda_b)=2\delta_{ab}$, and obey the same algebra as their single-particle counterparts.
\begin{equation}
  \hat\Lambda_a = \sum_{i=1}^{N}\hat\lambda_a^{(i)},
  \qquad
  [\hat\Lambda_a,\hat\Lambda_b] = 2i f_{abc}\hat\Lambda_c ,
  \qquad a = 1,\dots,8 .
  \label{eq:collective}
\end{equation}

We consider the following metrological protocol. A phase is imprinted onto the state by a linear generator $\hat H_{\mathbf n}=\mathbf n\cdot\hat{\boldsymbol\Lambda}$ and subsequently read out from a linear observable $\hat M_{\mathbf m}=\mathbf m\cdot\hat{\boldsymbol\Lambda}$.
The sensitivity of this protocol is then given by
\begin{equation}
  \chi^{-2}(\mathbf m,\mathbf n)
  =\frac{\big|\langle[\hat M_{\mathbf m},\hat H_{\mathbf n}]\rangle\big|^2}
         {(\Delta \hat M_{\mathbf m})^2}
  =\frac{\big(\mathbf m^{T}\mathbf C\,\mathbf n\big)^2}
         {\mathbf m^{T}\boldsymbol\Gamma\,\mathbf m} ,
  \label{eq:chi2}
\end{equation}
where we defined
\begin{align}
  \Gamma_{ab} &= \tfrac12\big\langle\{\hat\Lambda_a,\hat\Lambda_b\}\big\rangle
                 - \langle\hat\Lambda_a\rangle\langle\hat\Lambda_b\rangle ,
  \label{eq:Gamma}\\[2pt]
  C_{ab} &= -i\big\langle[\hat\Lambda_a,\hat\Lambda_b]\big\rangle
          = 2 f_{abc}\langle\hat\Lambda_c\rangle .
  \label{eq:C}
\end{align}

Maximizing the sensitivity in Eq.~\eqref{eq:chi2} over the read-out direction ($\mathbf{m}$) at a fixed generator has a well-known closed-form solution \cite{Gessner2017,Gessner2019}, $\mathbf m_{\rm opt}\propto\boldsymbol\Gamma^{-1}\mathbf C\,\mathbf n$. This results in 
\begin{equation}
  \chi^{-2}_{\rm opt}(\mathbf n)
  =\max_{\mathbf m}\chi^{-2}(\mathbf m,\mathbf n)
  =\mathbf n^{T}\mathbf M\,\mathbf n   
  \label{eq:M}
\end{equation}
with $\mathbf M = \mathbf C^{T}\boldsymbol\Gamma^{-1}\mathbf C $.

To quantify the metrological gain, this sensitivity must be normalized against the best possible separable state sensing the same generator \cite{Gessner2019}. For a linear generator, this bound is directly determined  by the spectral width of the single-particle generator $h_{\mathbf n}=\mathbf n\cdot\hat{\boldsymbol\lambda}$. Thus,
\begin{equation}
  F_{\rm SN}[\hat H_{\mathbf n}]
  =\max_{\hat\rho_{\rm cl}} F_Q[\hat\rho_{\rm cl},\hat H_{\mathbf n}]
  = N\,\delta h(\mathbf n)^2 ,
\end{equation}
with $\delta h(\mathbf n)=\lambda_{\max}(h_{\mathbf n})-\lambda_{\min}(h_{\mathbf n})$.

Therefore the optimal squeezing parameter measuring the metrological gain takes the form
\begin{equation}
  \xi^{2}_{\rm opt}
  =\min_{\mathbf n}
  \frac{F_{\rm SN}[\hat H_{\mathbf n}]}{\chi^{-2}_{\rm opt}(\mathbf n)}
  =\min_{\mathbf n}
  \frac{N\,\delta h(\mathbf n)^2}{\mathbf n^{T}\mathbf M\,\mathbf n}
  \label{eq:xi2opt}
\end{equation}
We note that Eq.~\eqref{eq:xi2opt} reduces to the familiar Wineland squeezing parameter~\cite{Wineland1992} if the generator and readout are restricted to an effective spin-1/2 $\mathfrak{su}(2)$ subalgebra, such as the bright-manifold operators $\hat{\boldsymbol B}$ of Appendix~\ref{app:bright_dark}. In that case $\delta h(\mathbf n)=1$ for all unit vectors $\mathbf n$, and the optimization reduces to the maximal eigenvalue of $\mathbf M$ within that subalgebra. For the full $\mathrm{SU}(3)$ operator set the ratio has to be optimized numerically over $\mathbf n$. For the bright state at $r=1$ the optimum lies in the bright-manifold subalgebra (Table~\ref{tab:measurement_ops_2}), and $\xi^2_{\rm opt}(t)$ coincides with the exact OAT result.

\section{Expanded Scaling Analysis of all-to-all Data}\label{app:scaling}
\begin{figure}[t]
    \centering
    \includegraphics[width=1\columnwidth]{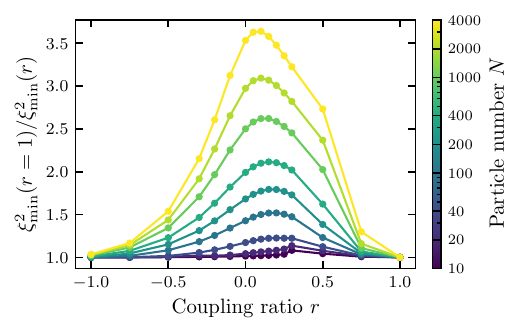}   
    \includegraphics[width=1\columnwidth]{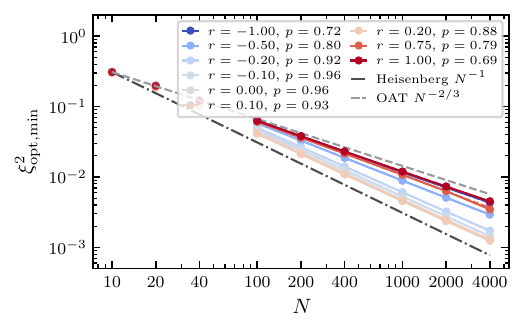}   
    \includegraphics[width=1\columnwidth]{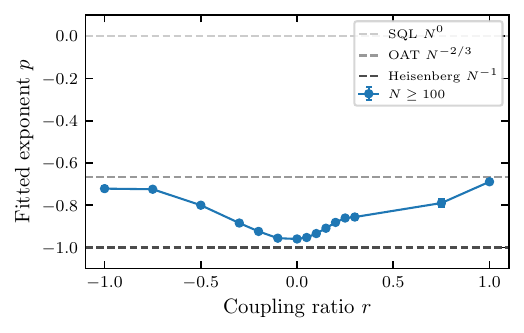}   
    \caption{Metrological advantage of three-level squeezing over OAT (realized at $r=1$) for all-to-all interactions ($\alpha=0$), from exact calculations with the optimal initial product state at each $r$ and $N$. Top: $ \xi^2_{\mathrm{min}}(r=1)/\xi^2_{\mathrm{min}}(r)$ quantifying the advantage of $r \neq 1$ compared to OAT realized at $r=1$. Middle: Minimal squeezing parameter $\xi^2_{\rm opt,min}$ versus system size for various values of $r$ (as indicated in legend). Fits performed over $N=100$ up to $N=4000$ to scaling form $\xi^2_{\rm{min}} = A \, N^{p} \, ( 1 +b N^{-1/3})$. Bottom: Extracted exponents $p$ versus the coupling ratio $r= J_2/J_1$ (error bars of fit exponent smaller than symbol size).
    \label{fig:SI_ED_xi_scaling}}
\end{figure}
We provide in Fig.~\ref{fig:SI_ED_xi_scaling} an expanded set of data for the behavior of the optimal squeezing parameter as a function of the coupling ratio $r$ and system size $N$. %
In the top panel we show the metrological advantage of operating in the full 3-level system at a ratio $r \neq  1$ by normalizing the achieved metrological gain by the one achieved at $r=1$, which realizes one-axis twisting (OAT) in the bright state manifold. Different curves correspond to different total particle number/system sizes $N$, and we observe a scalable/increasing advantage with $N$ of operating at non-unit coupling ratio $r$, maximized at $r=0.1$ for $N=4000$ among the simulated ratios. The relevance of all three levels for $r\neq1$ follows from the Hamiltonian rather than from the initial state. For $r\neq1$ the bright and dark manifolds are coupled [Eq.~\eqref{SI-eq:InteractionHamiltonianw_general_BD}], so the dynamics leaves any effective two-level subspace, and the optimal states for $r\neq\pm1$ populate all three levels (Tables~\ref{tab:optimal} and~\ref{tab:optimal_2}).

The middle panel shows the scaling of the raw minimal optimized squeezing parameter versus $N$, and includes a wider range of coupling ratios $r$ compared to Fig.~\ref{fig:ED_xi_vs_N} of the main text. We observe improved scaling behavior and a clear advantage over OAT ($r=1$) for any value of $r$. Finally, the bottom panel shows the extracted scaling exponent $p$ obtained from the fit to $\xi^2_{\rm{min}} = A \, N^{-p} \, ( 1 +b N^{-1/3})$ performed over the range $N=100$-$4000$ as a function of the coupling ratio $r$. Raising the lower limit of the fit window to $N=200$ ($400$) changes the exponents to $p=0.910(5)$ ($0.892(4)$) for $r=0.1$ and $p=0.677(2)$ ($0.671(1)$) for $r=1$, so that the difference $\Delta p=0.245$, $0.233$, $0.222$ is robust, while the window dependence exceeds the statistical fit errors, which therefore underestimate the systematic uncertainty. The maximal relative residuals of the fits over $100\le N\le4000$ are $1.4\%$ ($r=0.1$) and $0.5\%$ ($r=1$). For $r\approx-1$ the fitted $p>2/3$ reflects a change of the optimal state between $N=100$ and $200$; for $N\ge400$ we find $p\approx0.67$--$0.68$, close to OAT. %

\section{Optimal States and Generator/Measurement Operators \label{app:readout}}
In this section we provide the parameters of the optimal state for achieving the reported squeezing, ($p_1$, $p_2$, $p_3$), for the full range of Hamiltonians characterized by the coupling ratio $r = J_{2}/J_{1}$ (which we simulate for $r \in $ [-1.0,-0.75,-0.5,-0.3,-0.2,-0.1,0.0,0.05,0.1,0.15,0.2,0.25,0.3,0.5,0.75,1.0]) for a range of system sizes $N=10$ up to $N=4000$ (Table~\ref{tab:optimal}). %

We emphasize that the optimal state parameters depend on both the coupling ratio $r$ and system size $N$, and all squeezing values reported are for the optimal state. Since we consider the generalized squeezing parameter (Eq.~\eqref{eq:xi2opt}) the reported squeezing is the optimum over all initial product states $(\sqrt{p_1}\ket1+\sqrt{p_2}\ket2+\sqrt{p_3}\ket3)^{\otimes N}$ and all linear generator and read-out operators, for a given Hamiltonian and system size. Relative phases of the amplitudes do not affect $\xi^2_{\mathrm{opt}}$, since they can be removed by a product of single-site unitaries that commutes with $\hat H_{\mathrm{int}}$. 

Generally, both the optimal readout and generator depend on $r$ and $N$. With increasing system size, the optimal readout approaches a population observable
\begin{equation}
  \hat M_{\mathrm{opt}}\;\approx\;\hat N_2+k\,\hat S_z ,
\end{equation}
with $k\approx-0.28$. We quantify this by the weight $w_{\mathrm{pop}}=m_3^2+m_8^2$ of the normalized readout vector on the diagonal Gell-Mann operators [Eq.~\eqref{eq:SI_Gell-Mann}], which increases from $0.81$ at $N=10$ to $0.99$ at $N=4000$ (Tables~\ref{tab:optimal} and~\ref{tab:optimal_2}). Since $\hat N_1,\hat N_2,\hat N_3$ are conserved by $\hat H_{\mathrm{int}}$, the small remaining admixture of coherences is essential, in direct analogy to OAT, where the squeezed quadrature approaches the conserved twisting axis with an $N$-dependent tilt. Experimentally, the readout therefore corresponds to a small rotation followed by state-resolved detection of all three populations. Because it involves $\hat S_z=\hat N_1-\hat N_3$, which couples the bright and dark states, it has no counterpart within the effective two-level description.

For completeness we provide the full basis expansion of the optimal generator and readout operator at $N=4000$ and $r=0.1$ in Table~\ref{tab:measurement_ops} and for $r=1.0$ in Table~\ref{tab:measurement_ops_2}. We note that the results for $r=1.0$ are precisely the expected OAT dynamics in the $\hat{\boldsymbol B}
=\{\tfrac12\hat{S}_{x},\tfrac12\hat{Q}_{yz},\tfrac14(\hat{D}_{xy}+ \sqrt{3}\hat{Y})\}$ manifold for an initial $\hat{B}_x$ polarized state, with a $\hat{B}_z$ readout and $\hat{B}_y$ generator.

\begin{table}[t]
\centering
\footnotesize
\begin{tabular}{rrrrrrrrrr}
\hline
$N$ & $r^\ast$ & $p_1$ & $p_2$ & $p_3$ & $J N t_{\rm opt}/\hbar$ & $\xi^2_{\min}$ & $\xi^2_{\rm OAT}/\xi^2$ & $k$ & $w_{\rm pop}$ \\
\hline
10 & +0.30 & 0.04 & 0.41 & 0.56 & 6.3 & 0.2865 & 1.08 & -0.21 & 0.81 \\
20 & +0.30 & 0.05 & 0.42 & 0.53 & 7.9 & 0.1741 & 1.14 & -0.25 & 0.88 \\
40 & +0.20 & 0.23 & 0.36 & 0.41 & 7.9 & 0.0996 & 1.23 & -0.27 & 0.89 \\
100 & +0.20 & 0.27 & 0.38 & 0.36 & 11.6 & 0.0414 & 1.52 & -0.27 & 0.93 \\
200 & +0.15 & 0.32 & 0.40 & 0.28 & 14.5 & 0.0212 & 1.79 & -0.27 & 0.95 \\
400 & +0.15 & 0.34 & 0.42 & 0.24 & 18.8 & 0.0109 & 2.11 & -0.27 & 0.97 \\
1000 & +0.15 & 0.37 & 0.45 & 0.18 & 26.7 & 0.0046 & 2.62 & -0.27 & 0.98 \\
2000 & +0.10 & 0.41 & 0.46 & 0.13 & 34.7 & 0.0024 & 3.09 & -0.28 & 0.98 \\
4000 & +0.10 & 0.42 & 0.47 & 0.11 & 46.0 & 0.0012 & 3.64 & -0.28 & 0.99 \\
\hline
\end{tabular}
\caption{Optimal protocol at each system size. %
The initial state is the product state $\bigl(\sqrt{p_1}\lvert1\rangle+\sqrt{p_2}\lvert2\rangle+\sqrt{p_3}\lvert 3\rangle\bigr)^{\otimes N}$, with $\lvert1\rangle,\lvert2\rangle,\lvert3\rangle\equiv\lvert m=+1,0,-1\rangle$. We note that squeezing is independent of any relative phases in the initial state. %
$r^\ast$ is the coupling ratio maximizing the achievable squeezing $\xi_{\min}^2$ (out of the simulated ratios $r \in $ [-1.0,-0.75,-0.5,-0.3,-0.2,-0.1,0.0,0.05,0.1,0.15,0.2,0.25,0.3,0.5,0.75,1.0]), which occurs at time $t_{\mathrm{opt}}$, which we compare to the $\xi^2_{\mathrm{OAT}}$ result realized at $r=1$. %
$k$ and $w_{\rm pop}$ specify the read-out observable via  $\hat M^\ast \approx \hat N_2+k\,\hat S_z$, and $w_{\rm pop}$ quantifies the weight of $\hat M^\ast$ on the diagonal (population) generators (see Appendix~\ref{app:readout}).}
\label{tab:optimal}
\end{table}

\begin{table}[t]
\centering
\footnotesize
\begin{tabular}{rrrrrrrrrr}
\hline
$N$ & $r$ & $p_1$ & $p_2$ & $p_3$ & $J N t_{\rm opt}/\hbar$ & $\xi^2_{\min}$ & $\xi^2_{\rm OAT}/\xi^2$ & $k$ & $w_{\rm pop}$ \\
\hline
10 & +0.10 & 0.39 & 0.42 & 0.19 & 2.6 & 0.3039 & 1.02 & -0.30 & 0.77 \\
20 & +0.10 & 0.29 & 0.35 & 0.36 & 4.9 & 0.1860 & 1.06 & -0.29 & 0.85 \\
40 & +0.10 & 0.29 & 0.35 & 0.36 & 7.3 & 0.1007 & 1.21 & -0.30 & 0.87 \\
100 & +0.10 & 0.32 & 0.37 & 0.31 & 10.7 & 0.0420 & 1.50 & -0.29 & 0.92 \\
200 & +0.10 & 0.34 & 0.40 & 0.26 & 14.0 & 0.0214 & 1.77 & -0.29 & 0.94 \\
400 & +0.10 & 0.37 & 0.42 & 0.22 & 18.3 & 0.0110 & 2.10 & -0.29 & 0.96 \\
1000 & +0.10 & 0.39 & 0.44 & 0.16 & 26.3 & 0.0046 & 2.62 & -0.29 & 0.98 \\
2000 & +0.10 & 0.41 & 0.46 & 0.13 & 34.7 & 0.0024 & 3.09 & -0.28 & 0.98 \\
4000 & +0.10 & 0.42 & 0.47 & 0.11 & 46.0 & 0.0012 & 3.64 & -0.28 & 0.99 \\
\hline
\end{tabular}
\caption{Optimal state at fixed coupling ratio $r = 0.1$ for different system sizes $N$. %
The initial state is the product state $\bigl(\sqrt{p_1}\lvert1\rangle+\sqrt{p_2}\lvert2\rangle+\sqrt{p_3}\lvert 3\rangle\bigr)^{\otimes N}$, with $\lvert1\rangle,\lvert2\rangle,\lvert3\rangle\equiv\lvert m=+1,0,-1\rangle$. We note that squeezing is independent of any relative phases in the initial state. %
$r$ is the coupling ratio, $\xi_{\min}^2$ the achievable squeezing, which occurs at time $t_{\mathrm{opt}}$, which we compare to the $\xi^2_{\mathrm{OAT}}$ result realized at $r=1$. %
$k$ and $w_{\rm pop}$ specify the read-out observable via  $\hat M^\ast \approx \hat N_2+k\,\hat S_z$, and $w_{\rm pop}$ quantifies the weight of $\hat M^\ast$ on the diagonal (population) generators (see Appendix~\ref{app:readout}).}
\label{tab:optimal_2}
\end{table}

\begin{table}[t]
\centering
\small
\begin{tabular}{lrr|lrr}
\hline
\multicolumn{3}{c|}{dipole--quadrupole basis} & \multicolumn{3}{c}{Gell-Mann basis} \\
operator & $n^{\ast}$ & $m^{\ast}$ & $\Lambda_a$ & $n^{\ast}$ & $m^{\ast}$ \\
\hline
$S_x$ & +0.7426 & -0.0284 & $\Lambda_{1}$ & +0.8656 & +0.0102 \\
$S_y$ & +0.1414 & -0.0342 & $\Lambda_{2}$ & -0.0634 & -0.0013 \\
$S_z$ & -0.2031 & +0.4380 & $\Lambda_{3}$ & +0.0375 & +0.9926 \\
$Y$ & +0.1605 & +0.8933 & $\Lambda_{4}$ & -0.0556 & -0.0476 \\
$D_{xy}$ & -0.0556 & -0.0476 & $\Lambda_{5}$ & -0.2706 & +0.0558 \\
$Q_{xy}$ & -0.2706 & +0.0558 & $\Lambda_{6}$ & +0.1847 & -0.0504 \\
$Q_{yz}$ & -0.2311 & +0.0324 & $\Lambda_{7}$ & +0.2634 & -0.0470 \\
$Q_{xz}$ & +0.4815 & +0.0429 & $\Lambda_{8}$ & -0.2561 & -0.0673 \\
\hline
\end{tabular}
\caption{Optimal generator $n^{\ast}$ and readout $m^{\ast}$ at $N=4000$, $r=0.1$, expanded in two orthonormal bases of $\mathfrak{su}(3)$, the dipole-quadrupole and the Gell-Mann operators.}
\label{tab:measurement_ops}
\end{table}

\begin{table}[t]
\centering
\small
\begin{tabular}{lrr|lrr}
\hline
\multicolumn{3}{c|}{dipole--quadrupole basis} & \multicolumn{3}{c}{Gell-Mann basis} \\
operator & $n^{\ast}$ & $m^{\ast}$ & $\Lambda_a$ & $n^{\ast}$ & $m^{\ast}$ \\
\hline
$S_x$ & -0.0000 & +0.0000 & $\Lambda_{1}$ & -0.0000 & -0.0000 \\
$S_y$ & +0.0000 & -0.0000 & $\Lambda_{2}$ & +0.7060 & +0.0395 \\
$S_z$ & +0.0000 & +0.0000 & $\Lambda_{3}$ & -0.0419 & +0.7488 \\
$Y$ & -0.0484 & +0.8647 & $\Lambda_{4}$ & -0.0279 & +0.4992 \\
$D_{xy}$ & -0.0279 & +0.4992 & $\Lambda_{5}$ & -0.0000 & -0.0000 \\
$Q_{xy}$ & -0.0000 & -0.0000 & $\Lambda_{6}$ & -0.0000 & +0.0000 \\
$Q_{yz}$ & +0.9984 & +0.0558 & $\Lambda_{7}$ & -0.7060 & -0.0395 \\
$Q_{xz}$ & +0.0000 & -0.0000 & $\Lambda_{8}$ & +0.0242 & -0.4323 \\
\hline
\end{tabular}
\caption{Optimal generator $n^{\ast}$ and readout $m^{\ast}$ at $N=4000$, $r=1.0$, expanded in two orthonormal bases of $\mathfrak{su}(3)$, the dipole-quadrupole and the Gell-Mann operators.}
\label{tab:measurement_ops_2}
\end{table}

\clearpage

\section{Extended data on the optimal state for dipolar interactions}\label{app:dipolar_optimal_state}
\label{sec:sm_mps}
\newcommand{\smGNine}{0.274}
\newcommand{\smGsigNine}{0.000}
\newcommand{\smGtopNine}{0.274}
\newcommand{\smRstarNine}{0.348}
\newcommand{\smGTwelve}{0.339}
\newcommand{\smGsigTwelve}{0.000}
\newcommand{\smGtopTwelve}{0.339}
\newcommand{\smRstarTwelve}{0.363}
\newcommand{\smGSixteen}{0.419}
\newcommand{\smGsigSixteen}{0.000}
\newcommand{\smGtopSixteen}{0.419}
\newcommand{\smRstarSixteen}{0.385}
\newcommand{\smGTwenty}{0.476}
\newcommand{\smGsigTwenty}{0.002}
\newcommand{\smGtopTwenty}{0.475}
\newcommand{\smRstarTwenty}{0.404}
\newcommand{\smGTwentyFive}{0.535}
\newcommand{\smGsigTwentyFive}{0.004}
\newcommand{\smGtopTwentyFive}{0.534}
\newcommand{\smRstarTwentyFive}{0.422}
\newcommand{\smGThirty}{0.574}
\newcommand{\smGsigThirty}{0.002}
\newcommand{\smGtopThirty}{0.572}
\newcommand{\smRstarThirty}{0.440}
\newcommand{\smGThirtySix}{0.612}
\newcommand{\smGsigThirtySix}{0.009}
\newcommand{\smGtopThirtySix}{0.603}
\newcommand{\smRstarThirtySix}{0.457}
\newcommand{\smGFortyTwo}{0.640}
\newcommand{\smGsigFortyTwo}{0.011}
\newcommand{\smGtopFortyTwo}{0.628}
\newcommand{\smRstarFortyTwo}{0.470}
\newcommand{\smGFortyNine}{0.661}
\newcommand{\smGsigFortyNine}{0.027}
\newcommand{\smGtopFortyNine}{0.627}
\newcommand{\smRstarFortyNine}{0.481}
\newcommand{\smGlbFortyNine}{0.655}
\newcommand{\smGlbsigFortyNine}{0.028}
\newcommand{\smGlbtopFortyNine}{0.624}
\newcommand{\smRhoMinB}{0.16}
\newcommand{\smRhoMaxB}{0.42}
\newcommand{\smRhoMinXY}{0.14}
\newcommand{\smRhoMaxXY}{0.27}
\newcommand{\smNrhoB}{15}
\newcommand{\smRhoX}{0.35}
\newcommand{\smRhoMin}{0.14}
\newcommand{\smRhoMax}{0.42}
\newcommand{\smNcells}{24}
\newcommand{\smNruns}{79}
\newcommand{\smTailMax}{0.039}
\newcommand{\smRhoSens}{0.42}
\newcommand{\smGsensMax}{0.013}
\newcommand{\smSigmaDt}{2\times10^{-4}}
\newcommand{\smLogiC}{4.065}
\newcommand{\smLogiD}{2.760}
\newcommand{\smNdtwa}{18}
\newcommand{\smDtwaNmax}{10^{4}}
\newcommand{\smDtwaRFortyTwo}{0.477}
\newcommand{\smDtwaRFortyNine}{0.484}
\newcommand{\smDtwaNextrapMin}{49}
\newcommand{\smDtwaNedMax}{16}
\newcommand{\smDtwaNmpsMin}{20}
\newcommand{\smDtwaNmpsMax}{42}

In this section we provide additional details on the numerical simulations underlying Fig.~\ref{fig:dipolar_results}, in particular the extrapolation of the MPS results in bond dimension and the precise initial states of the dTWA simulations.

For the $3\times3$, $3\times4$ and $4\times4$ lattices ($N=9$, 12 and 16) we compute the time evolution by exact diagonalization (ED). %
For $20\le N\le49$ ($5\times4$ to $7\times7$ lattices) we use matrix-product-state (MPS) time evolution with the two-site time-dependent variational principle (TDVP) \cite{Haegeman2011,Haegeman2016} as implemented in \textsc{ITensors.jl} \cite{Itensor1,Itensor2}, with time step $dt=0.05\,\hbar/J$, truncation cutoff $10^{-10}$ and bond dimension $\chi\le512$. All $\xi^2$ in this section are given in dB.

The initial states are product states parametrized by $(q_0,s)$ as
\begin{equation}
  \begin{gathered}
    |\psi_c(0)\rangle=\textstyle\bigotimes_i\bigl(c_1|1\rangle+c_2|2\rangle+c_3|3\rangle\bigr),\\
    c_2=\sqrt{q_0},\qquad c_{1,3}=\sqrt{\tfrac12(1-q_0)(1\pm s)}.
  \end{gathered}
  \label{eq:sm_state}
\end{equation}

We minimize $\xi^2$ over $(q_0,s)$ by a local search and over $t$ by a parabolic fit around its minimum. The bright reference $\xi^2_{\mathrm{bright}}$ is computed from the XY state $(q_0,s)=(\frac12,+1)$, which has the same squeezing as the bright state $(q_0,s)=(\frac12,0)$ at $r=1$ and, since it does not populate $|3\rangle$, evolves exactly as the spin-$\frac12$ dipolar XY model for any $r$. It is computed by MPS for $N\ge30$ and by exact diagonalization of the spin-$\frac12$ XY model for $N=20$ and 25.

\begin{figure}[!tb]
\includegraphics[width=\columnwidth]{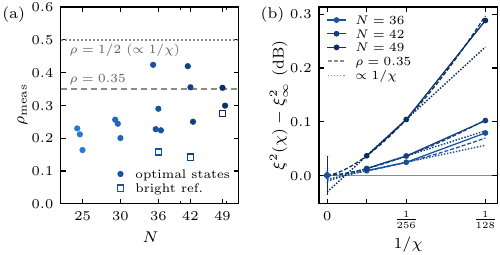}
\caption{Extrapolation in bond dimension. (a) Measured ratio $\rho_{\mathrm{meas}}$ for the optimal-state points (circles; darker for larger $N$, offset slightly in $N$ by $r$) and the bright references (squares). Dashed: $\rho=\smRhoX$, Eq.~\eqref{eq:sm_mps_tail}; dotted: $\rho=\frac12$ (linear in $1/\chi$). %
(b) $\xi^2(\chi)-\xi^2_\infty$ against $1/\chi$ at $r=0.45$ for $N=36$ ($6\times6$), 42 ($6\times7$) and 49 ($7\times7$): computed values (circles, solid lines), the extrapolation of Eq.~\eqref{eq:sm_mps_tail} (dashed) and a straight line in $1/\chi$ (dotted) through the two largest $\chi$, and $\xi^2_\infty$ with its error bar (diamonds).}
\label{fig:sm_mps_chi}
\end{figure}

Every MPS value, including the bright state reference, is extrapolated to $\chi\to\infty$ from the two largest computed bond dimensions,
\begin{equation}
  \xi^2_\infty=\xi^2(\chi_{\max})+\Delta\,\frac{\rho}{1-\rho},
  \label{eq:sm_mps_tail}
\end{equation}
with $\Delta=\xi^2(\chi_{\max})-\xi^2(\chi_{\max}/2)$ and $\rho=\smRhoX$, i.e.\ assuming a constant ratio $\rho$ of successive differences per doubling of $\chi$.
The measured ratios $\rho_{\mathrm{meas}}=\Delta(256,512)/\Delta(128,256)$ [Fig.~\ref{fig:sm_mps_chi}(a) and Table~\ref{tab:sm_mps_ladders}] lie between \smRhoMin{} and \smRhoMax{}, which excludes a linear extrapolation in $1/\chi$ ($\rho=\frac12$).

Figure~\ref{fig:sm_mps_chi}(b) shows the resulting extrapolation with $\rho=\smRhoX$ at $r=0.45$ (dashed line), compared with a linear extrapolation in $1/\chi$ (dotted). For all systems in Table~\ref{tab:sm_mps_ladders}, $\xi^2_{\mathrm{opt}}$ decreases and $\xi^2_{\mathrm{bright}}$ increases with $\chi$, so $G$ grows with $\chi$. Thus, our choice of $\rho$ gives a smaller, i.e.\ more conservative, increase of $G$ than an extrapolation in $1/\chi$. It is consistent with other common extrapolation schemes: extrapolation with the measured ratio $\rho$ for each state, a straight line in the cumulative discarded weight, and a parabolic fit in $1/\chi$ through $\chi=128$, 256 and 512 change $G_{\mathrm{dB}}(r^*)$ by at most 0.01\,dB at $N=36$, 42 and 49,  within the error bars of Table~\ref{tab:sm_gstar}. For $4\times4$, where the exact result is known, Eq.~\eqref{eq:sm_mps_tail} applied to bond dimensions $\chi=32$, 64 and 128 reproduces it to within 0.006\,dB.

The error bar $\sigma$ combines in quadrature the truncation error $\sigma_{\mathrm{trunc}}=|\Delta|\rho/(1-\rho)$, i.e.\ the full size of the extrapolation (with $\rho=\frac12$ for the MPS bright references), the time-step error $\sigma_{dt}$ given by the change of $\xi^2$ when $dt$ is halved, and interpolation errors. For $N\ge42$, $\sigma\simeq\sigma_{\mathrm{trunc}}$. The error bars of the MPS points in Fig.~\ref{fig:dipolar_results}(a) combine $\sigma$ of the optimal state and of the bright reference in quadrature.

For each $N\ge20$, $r^*$ and $\xi^2_{\mathrm{opt}}(r^*)$ are obtained from the maximum of a parabolic fit through the three extrapolated values around the most squeezed one. The error of $G(r^*)$ adds in quadrature the resulting uncertainty of the optimal state value $\sigma_{\mathrm{opt}}$ and $\sigma_{\mathrm{bright}}$. Table~\ref{tab:sm_gstar} lists
$G_{\mathrm{dB}}(r^*)=\xi^2_{\mathrm{bright}}-\xi^2_{\mathrm{opt}}(r^*)$.

The dTWA points of Fig.~\ref{fig:dipolar_results}(b) use $10^4$ trajectories per lattice and start from the product states of Table~\ref{tab:sm_dtwa_states}, one per lattice, at $r=r^*(N)$. For $N\ge\smDtwaNextrapMin$, $r^*$ is extrapolated as $r^*(N)=1/\{1+\exp[-(\ln N-c)/d]\}$ with $c=\smLogiC$ and $d=\smLogiD$, fitted to $r^*$ for $16\le N\le 42$.

\begin{table*}
\caption{MPS results (dB) for the optimal state at ratio $r$ and for the bright reference (XY). Optimum $\xi^2_\chi$ at bond
dimension $\chi$, $\Delta$, $\rho_{\mathrm{meas}}$, and the extrapolated value $\xi^2_\infty$ [Eq.~\eqref{eq:sm_mps_tail};
ED: exact diagonalization] plotted in Fig.~\ref{fig:dipolar_results}, with its error $\sigma$.}
\label{tab:sm_mps_ladders}
\footnotesize
\begin{ruledtabular}
\begin{tabular}{llrrrrrrr}
$N$ & $r$ & \multicolumn{1}{c}{$\xi^2_{128}$} & \multicolumn{1}{c}{$\xi^2_{256}$} & \multicolumn{1}{c}{$\xi^2_{512}$} & \multicolumn{1}{c}{$\Delta$} & \multicolumn{1}{c}{$\rho_{\mathrm{meas}}$} & \multicolumn{1}{c}{$\xi^2_{\infty}$} & \multicolumn{1}{c}{$\sigma$} \\
\hline
20 ($5\times4$) & 0.30 & $-5.5108$ & $-5.5147$ & -- & $-0.0039$ & -- & $-5.5168$ & 0.0030 \\
 & 0.35 & $-5.6295$ & $-5.6324$ & -- & $-0.0029$ & -- & $-5.6340$ & 0.0027 \\
 & 0.40 & $-5.6651$ & $-5.6671$ & -- & $-0.0020$ & -- & $-5.6681$ & 0.0022 \\
 & 0.45 & $-5.6427$ & $-5.6434$ & -- & $-0.0007$ & -- & $-5.6438$ & 0.0018 \\
 & 0.50 & $-5.5857$ & $-5.5857$ & -- & $-5\times10^{-5}$ & -- & $-5.5858$ & 0.0016 \\
 & XY & $-5.1922$ & $-5.1922$ & -- & $5\times10^{-6}$ & -- & $-5.1923^{\mathrm{ED}}$ & 0.0002 \\
\hline
25 ($5\times5$) & 0.30 & $-5.9326$ & $-5.9535$ & -- & $-0.0208$ & -- & $-5.9647$ & 0.0163 \\
 & 0.35 & $-6.0999$ & $-6.1179$ & $-6.1220$ & $-0.0041$ & 0.23 & $-6.1242$ & 0.0041 \\
 & 0.40 & $-6.1749$ & $-6.1888$ & $-6.1917$ & $-0.0029$ & 0.21 & $-6.1933$ & 0.0049 \\
 & 0.45 & $-6.1776$ & $-6.1871$ & $-6.1887$ & $-0.0016$ & 0.16 & $-6.1895$ & 0.0009 \\
 & 0.50 & $-6.1309$ & $-6.1366$ & -- & $-0.0057$ & -- & $-6.1397$ & 0.0035 \\
 & XY & $-5.6656$ & $-5.6651$ & -- & $0.0005$ & -- & $-5.6652^{\mathrm{ED}}$ & $<10^{-4}$ \\
\hline
30 ($5\times6$) & 0.35 & $-6.4691$ & $-6.4964$ & $-6.5034$ & $-0.0070$ & 0.26 & $-6.5072$ & 0.0072 \\
 & 0.40 & $-6.5803$ & $-6.6004$ & $-6.6053$ & $-0.0049$ & 0.24 & $-6.6079$ & 0.0056 \\
 & 0.45 & $-6.6051$ & $-6.6204$ & $-6.6234$ & $-0.0031$ & 0.20 & $-6.6251$ & 0.0017 \\
 & 0.50 & $-6.5710$ & $-6.5805$ & -- & $-0.0096$ & -- & $-6.5857$ & 0.0054 \\
 & XY & $-6.0538$ & $-6.0526$ & -- & $0.0011$ & -- & $-6.0520$ & 0.0011 \\
\hline
36 ($6\times6$) & 0.35 & $-6.8253$ & $-6.8909$ & $-6.9187$ & $-0.0278$ & 0.42 & $-6.9336$ & 0.0150 \\
 & 0.40 & $-6.9992$ & $-7.0568$ & $-7.0699$ & $-0.0131$ & 0.23 & $-7.0769$ & 0.0291 \\
 & 0.45 & $-7.0436$ & $-7.0981$ & $-7.1139$ & $-0.0158$ & 0.29 & $-7.1224$ & 0.0085 \\
 & 0.50 & $-7.0395$ & $-7.0820$ & $-7.0917$ & $-0.0097$ & 0.22 & $-7.0969$ & 0.0052 \\
 & XY & $-6.5206$ & $-6.5127$ & $-6.5115$ & $0.0012$ & 0.16 & $-6.5108$ & 0.0012 \\
\hline
42 ($6\times7$) & 0.40 & $-7.3400$ & $-7.4078$ & $-7.4362$ & $-0.0284$ & 0.42 & $-7.4515$ & 0.0153 \\
 & 0.45 & $-7.4250$ & $-7.4909$ & $-7.5143$ & $-0.0234$ & 0.36 & $-7.5270$ & 0.0126 \\
 & 0.50 & $-7.4329$ & $-7.4946$ & $-7.5101$ & $-0.0154$ & 0.25 & $-7.5184$ & 0.0083 \\
 & XY & $-6.9084$ & $-6.8963$ & $-6.8946$ & $0.0017$ & 0.14 & $-6.8936$ & 0.0017 \\
\hline
49 ($7\times7$) & 0.40 & $-7.5499$ & $-7.7499$ & $-7.8219$ & $-0.0727$ & -- & $-7.8610$ & 0.0391 \\
 & 0.45 & $-7.6855$ & $-7.8697$ & $-7.9373$ & $-0.0675$ & 0.35 & $-7.9736$ & 0.0364 \\
 & 0.50 & $-7.7440$ & $-7.9084$ & $-7.9595$ & $-0.0502$ & 0.30 & $-7.9866$ & 0.0270 \\
 & XY & $-7.3692$ & $-7.3430$ & $-7.3358$ & $0.0072$ & 0.27 & $-7.3319$ & 0.0072 \\
\end{tabular}

\end{ruledtabular}
\end{table*}

\begin{table}[!htb]
\caption{Maximal gain (dB) from the extrapolated values. $G^{\mathrm{top}}_{\mathrm{dB}}$ from the values at the largest computed $\chi$. ED: exact diagonalization.}
\label{tab:sm_gstar}
\footnotesize
\begin{ruledtabular}
\begin{tabular}{lrrrrr}
$N$ & \multicolumn{1}{c}{$r^*$} & \multicolumn{1}{c}{$\xi^2_{\mathrm{bright}}$} & \multicolumn{1}{c}{$\xi^2_{\mathrm{opt}}(r^*)$} & \multicolumn{1}{c}{$G_{\mathrm{dB}}(r^*)$} & \multicolumn{1}{c}{$G^{\mathrm{top}}_{\mathrm{dB}}$} \\
\hline
20 ($5\times4$) & $0.404(4)$ & $-5.1923(2)^{\mathrm{ED}}$ & $-5.6683$ & $0.476(2)$ & $0.475$ \\
25 ($5\times5$) & $0.422(3)$ & $-5.6652^{\mathrm{ED}}$ & $-6.2006$ & $0.535(4)$ & $0.534$ \\
30 ($5\times6$) & $0.440(5)$ & $-6.0520(11)$ & $-6.6262$ & $0.574(2)$ & $0.572$ \\
36 ($6\times6$) & $0.457(9)$ & $-6.5108(12)$ & $-7.1231$ & $0.612(9)$ & $0.603$ \\
42 ($6\times7$) & $0.470(8)$ & $-6.8936(17)$ & $-7.5336$ & $0.640(11)$ & $0.628$ \\
49 ($7\times7$) & $0.481(28)$ & $-7.3319(72)$ & $-7.9934$ & $0.661(27)$ & $0.627$ \\
\end{tabular}

\end{ruledtabular}
\end{table}

\begin{table}[!htb]
\caption{Parameters for the dTWA results in Fig.~\ref{fig:dipolar_results}(b). Interaction ratio $r$ and state parameters $(q_0,s)$ of Eq.~\eqref{eq:sm_state}.
For $N\le\smDtwaNedMax$, the exact optimum (ED), for $\smDtwaNmpsMin\le N\le\smDtwaNmpsMax$, $r$ is the MPS estimate of $r^*$ for $N\ge\smDtwaNextrapMin$, $r$ is the extrapolated $r^*(N)$.}
\label{tab:sm_dtwa_states}
\footnotesize
\begin{ruledtabular}
\begin{tabular}{lrrr}
$N$ & \multicolumn{1}{c}{$r$} & \multicolumn{1}{c}{$q_0$} & \multicolumn{1}{c}{$s$} \\
\hline
9 ($3\times3$) & $0.3478$ & $0.40353$ & $-0.82120$ \\
12 ($3\times4$) & $0.3631$ & $0.41306$ & $-0.79841$ \\
16 ($4\times4$) & $0.3852$ & $0.42490$ & $-0.76330$ \\
\hline
20 ($5\times4$) & $0.4047$ & $0.42930$ & $-0.72880$ \\
25 ($5\times5$) & $0.4229$ & $0.43430$ & $-0.69710$ \\
30 ($5\times6$) & $0.4399$ & $0.43910$ & $-0.66880$ \\
36 ($6\times6$) & $0.4582$ & $0.44440$ & $-0.63930$ \\
42 ($6\times7$) & $0.4770$ & $0.44950$ & $-0.61000$ \\
\hline
49 ($7\times7$) & $0.4843$ & $0.45210$ & $-0.59870$ \\
64 ($8\times8$) & $0.5085$ & $0.45860$ & $-0.56230$ \\
100 ($10\times10$) & $0.5488$ & $0.46820$ & $-0.50550$ \\
196 ($14\times14$) & $0.6082$ & $0.48010$ & $-0.42300$ \\
400 ($20\times20$) & $0.6678$ & $0.48810$ & $-0.34760$ \\
784 ($28\times28$) & $0.7195$ & $0.49320$ & $-0.28470$ \\
1600 ($40\times40$) & $0.7686$ & $0.49620$ & $-0.22910$ \\
3136 ($56\times56$) & $0.8091$ & $0.49810$ & $-0.18420$ \\
6400 ($80\times80$) & $0.8459$ & $0.49900$ & $-0.14680$ \\
10000 ($100\times100$) & $0.8658$ & $0.49940$ & $-0.12650$ \\
\end{tabular}

\end{ruledtabular}
\end{table}

\clearpage

\section{Numerical Benchmarks} \label{app:benchmarks}
In this section we benchmark the generalized discrete truncated Wigner approximation (dTWA)~\cite{Schachenmayer2015,Zhu_NewJournalofPhysics_21_2019}, a semiclassical method widely used to study the dynamics of interacting quantum spin systems, for the Hamiltonian of the main text and a representative initial state. We first study the convergence of the dTWA results with the number of sampled trajectories, and then compare them against exact calculations and, for larger dipolar lattices, against matrix product state (MPS) simulations.

Throughout, we consider the coupling ratios $r=0.1$, $0.2$, $0.4$ and $1$, and keep the initial state fixed to $\ket{\Psi_0}=\bigotimes_{i=1}^N\big(\ket{1}+\sqrt{2}\,\ket{2}+\ket{3}\big)_i/2$, the spin coherent state fully polarized along the $+x$ axis, i.e., the eigenstate of $S_x$ with maximum eigenvalue. We denote by $\xi^2_{\mathrm{opt}}(t)$ the squeezing parameter optimized over generator and readout as in the main text, by $\xi^2_{\min}=\min_t\xi^2_{\mathrm{opt}}(t)$ its minimum, and by $t_{\mathrm{opt}}$ the time at which the minimum is reached. Unless stated otherwise, dTWA results use $N_{\rm traj}=4\times10^4$ trajectories.

\subsection{Convergence of Truncated Wigner Simulations}

\begin{figure}[!tbh]
    \centering
    \includegraphics[width=\columnwidth]{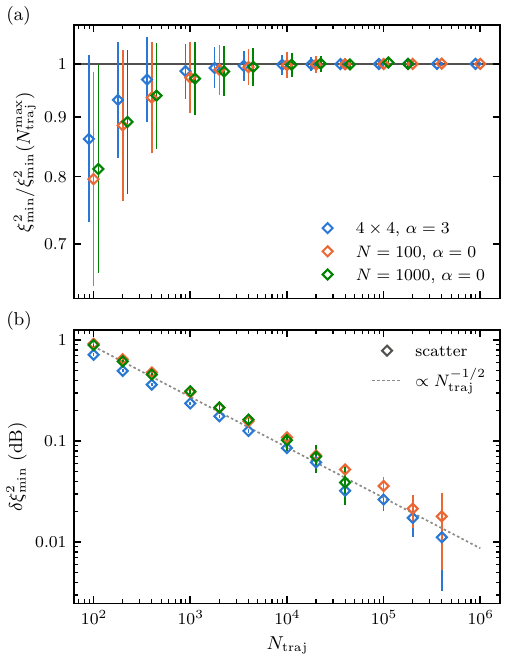}
    \caption{Convergence of dTWA in the number of trajectories $N_{\rm traj}$ for $r=0.4$ and the initial state $\bigotimes_i(|1\rangle+\sqrt{2}\,|2\rangle+|3\rangle)_i/2$, for a $4\times4$ lattice with dipolar interactions and for all-to-all interactions with $N=100$ and $N=1000$.
  (a)~Mean of $\xi^2_{\min}$ over the sub-ensembles of $N_{\rm traj}$ trajectories, relative to its value for $N_{\rm traj}^{\max}$. Error bars are standard deviation over the sub-ensembles, i.e., the error of a single sub-ensemble (jackknife error where only one sub-ensemble exists).
  The mean approaches the converged value from below, with a deviation that decreases as $1/N_{\rm traj}$.
  (b)~$1\sigma$ statistical error of
  $\xi^2_{\min}$. Standard deviation over the sub-ensembles, with its uncertainty. The error decreases as $N_{\rm traj}^{-1/2}$ for all three systems.
    \label{fig:SI_benchmark_convergence}}
\end{figure}

Before comparing dTWA against exact methods, we establish the statistical convergence of the Monte Carlo sampling with respect to the number of trajectories $N_{\mathrm{traj}}$.

Specifically, we consider $r=0.4$ for a $4\times 4$ square lattice with dipolar interactions and for all-to-all interactions with $N=100$ and $N=1000$, and analyze the convergence of $\xi^2_{\min}$, our main quantity of interest. For each system, one sequence of $N_{\rm traj}^{\max}=10^6$  ($4\times4$, $N=100$) or $1.6\times10^5$ ($N=1000$) trajectories is divided into disjoint sub-ensembles of $N_{\rm traj}$ trajectories (at most 400 per $N_{\rm traj}$), which are then analyzed independently.

Because $\xi^2_{\mathrm{opt}}$ is obtained by minimizing a nonlinear ratio of sample covariances and commutators over the generator direction $\mathbf{n}$, and $\xi^2_{\min}$ additionally by minimizing over time, finite-sample noise introduces a systematic downward bias at small $N_{\mathrm{traj}}$. This bias decays as $1/N_{\mathrm{traj}}$ [Fig.~\ref{fig:SI_benchmark_convergence}(a)], from $14\text{--}20\%$ ($0.6\text{--}1.0\,\mathrm{dB}$) at $N_{\mathrm{traj}}=100$ to below $1.5\%$ for $N_{\mathrm{traj}} \ge 2\times10^3$ and at most $0.1\%$ ($0.004\,$dB) at $N_{\mathrm{traj}}=4\times10^4$.

The $1\sigma$ statistical error of $\xi^2_{\min}$ scales as $N_{\mathrm{traj}}^{-1/2}$ [Fig.~\ref{fig:SI_benchmark_convergence}(b)], $\delta\xi^2_{\min}\approx(8\text{-}11)\,\mathrm{dB}/\sqrt{N_{\mathrm{traj}}}$, i.e., $0.24\text{--}0.31\,$dB for $10^3$, about $0.1\,$dB for $10^4$, and $0.03\text{--}0.05\,$dB for the $4\times10^4$ trajectories used below, where the bias is therefore negligible.

We show the statistical errors of dTWA in the benchmarks below. For the chosen number of trajectories they are typically smaller than the symbol size or line width, and in most cases several times smaller than the systematic deviations from the exact results.

\subsection{Benchmarking Truncated Wigner Calculations against Exact and MPS Calculations}

\subsubsection{All-to-all interactions}

\begin{figure}[!tb]
    \centering
    \includegraphics[width=\columnwidth]{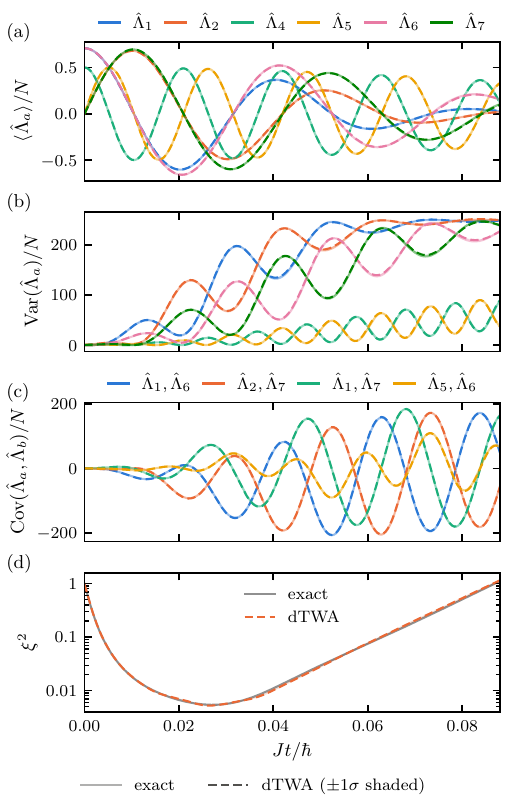}
    \caption{dTWA versus exact dynamics for all-to-all interactions ($\alpha=0$), $r=0.4$ and $N=1000$,  with initial state $\bigotimes_i(|1\rangle+\sqrt{2}\,|2\rangle+|3\rangle)_i/2$.
  (a)~Expectation values $\langle\hat\Lambda_a\rangle/N$ of the collective Gell-Mann operators $\hat\Lambda_a$ (colors, key above), (b)~their variances (same colors),  and (c)~the four largest distinct off-diagonal covariances, with  $\mathrm{Cov}(\hat\Lambda_a,\hat\Lambda_b)=\frac12\langle\{\hat\Lambda_a,\hat\Lambda_b\}\rangle
  -\langle\hat\Lambda_a\rangle\langle\hat\Lambda_b\rangle$; $\langle\hat\Lambda_3\rangle$, $\langle\hat\Lambda_8\rangle$ and the (co)variances among $\hat\Lambda_3$ and $\hat\Lambda_8$ are conserved and not shown. (d)~Squeezing parameter $\xi^2_{\mathrm{opt}}$. Exact solution (solid lines) and dTWA (dashed lines) with its $\pm1\sigma$
  statistical uncertainty shaded ($4\times10^4$ trajectories). 
    \label{fig:SI_benchmark_alpha0_traces}}
\end{figure}

\begin{figure}[!tb]
    \centering
    \includegraphics[width=\columnwidth]{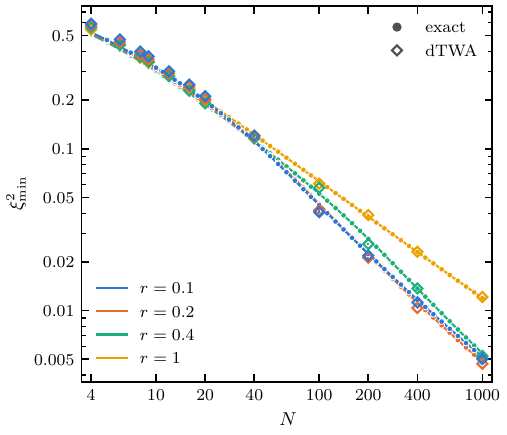}
    \caption{Minimal squeezing $\xi^2_{\min}$ versus $N$ for all-to-all interactions ($\alpha=0$) and $r=0.1$, $0.2$, $0.4$, $1$ (colors), with the initial state of Fig.~\ref{fig:SI_benchmark_alpha0_traces}. Exact results: small filled circles on solid lines; dTWA: open diamonds ($4\times10^4$ trajectories), with $1\sigma$ errors smaller than the symbols.\label{fig:SI_benchmark_alpha_0_scaling}}
\end{figure}

For all-to-all interactions, the dynamics remains in the permutationally symmetric subspace, which can be described by three bosonic modes with Fock states $\ket{n_1,n_2,n_3}$, $n_1+n_2+n_3=N$, spanning a space of dimension $(N+1)(N+2)/2$. This allows us to obtain the exact dynamics for large systems (up to $N=4000$ in the main text; here we benchmark $N\le1000$). Here, times are given in units of $\hbar/J$, with $J$ the coupling of every pair.

Figure~\ref{fig:SI_benchmark_alpha0_traces} compares the time evolution of the collective expectation values $\langle\hat{\Lambda}_a\rangle/N$, their variances, the dominant off-diagonal covariances, and the resulting $\xi^2_{\mathrm{opt}}(t)$ between the exact solution and dTWA for $N=1000$ and $r=0.4$. We find excellent quantitative agreement for all one- and two-point observables up to and beyond the optimal squeezing time. The relative error of the full covariance matrix, $\|\mathrm{Cov}_{\rm dTWA}-\mathrm{Cov}_{\rm exact}\|/\|\mathrm{Cov}_{\rm exact}\|$ (Frobenius norm), stays below $1.5\%$, close to its statistical uncertainty of about $1\%$. dTWA reaches $\xi^2_{\min}=-22.80(5)\,$dB, compared with the exact $-22.65\,$dB, both at $Jt_{\mathrm{opt}}/\hbar\approx0.027$.

Figure~\ref{fig:SI_benchmark_alpha_0_scaling} further benchmarks $\xi^2_{\min}$ as a function of system size, $4\le N\le1000$, for all four coupling ratios. dTWA accurately captures both the magnitude of $\xi^2_{\min}$ and its scaling with system size for all $r$, with the exact (filled markers) and dTWA (open markers) results largely overlapping.  For $N\le40$, dTWA slightly underestimates the squeezing, by at most $0.53\,$dB at $N=4$ (at $r=1$ it agrees to within $0.03\,$dB at $N=20$ and $40$). For $N\ge100$, the deviations are of either sign and at most $0.44\,$dB, and at $N=1000$ dTWA agrees with the exact result to within $0.16\,$dB for all $r$.

\clearpage
\subsubsection{Dipolar interactions}

\begin{figure}[!tb]
    \centering
    \includegraphics[width=\columnwidth]{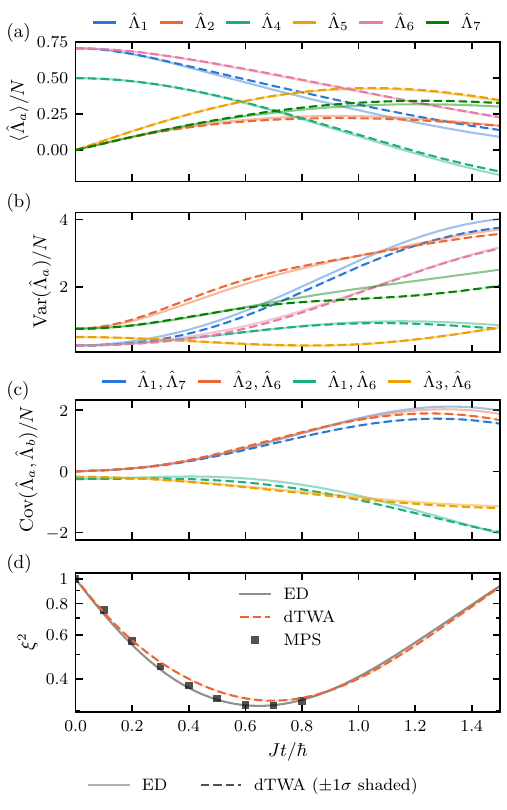}
    \caption{As Fig.~\ref{fig:SI_benchmark_alpha0_traces} for dipolar interactions on a $4\times4$ square lattice with open boundaries and $r=0.4$; times in units of $\hbar/J$ with $J$ the nearest-neighbor coupling. The reference (solid lines) is exact diagonalization (ED). Squares in (d): MPS (two-site TDVP, time step $0.05\,\hbar/J$, bond dimension 256; every second time step shown).
    \label{fig:SI_benchmark_alpha3_traces}}
\end{figure}

\begin{figure}[!tb]
    \centering
    \includegraphics[width=\columnwidth]{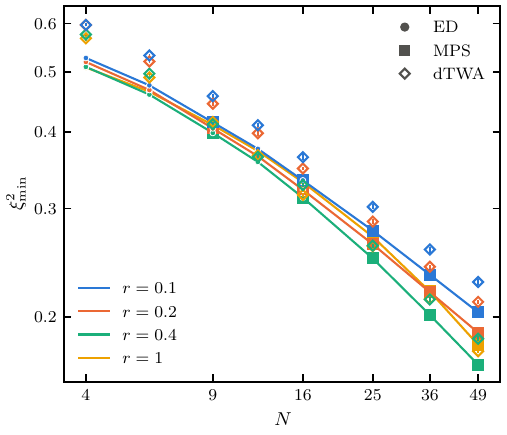}
    \caption{Minimal squeezing $\xi^2_{\min}$ versus $N$ for dipolar interactions ($\alpha=3$) on lattices with open boundaries ($2\times2$, $2\times3$, $3\times3$, $3\times4$, and $L\times L$ up to $7\times7$) and $r=0.1$, $0.2$, $0.4$, $1$ (colors), with the initial state of Fig.~\ref{fig:SI_benchmark_alpha3_traces}. Small filled circles: ED ($N\le16$); squares: MPS ($3\times3$ to $7\times7$, as in Fig.~\ref{fig:SI_benchmark_alpha3_traces}); solid lines connect the reference values (ED where available). Open diamonds: dTWA, with $1\sigma$ errors smaller than the symbols.
    \label{fig:SI_benchmark_alpha3_scaling}}
\end{figure}

In contrast, for dipolar interactions, $J_{ij}=J/|\mathbf r_i-\mathbf r_j|^3$ with $J$ the nearest-neighbor coupling, the dynamics is no longer restricted to the permutationally symmetric subspace, and exact calculations are computationally more limited. For $N \le 16$ (Hilbert-space dimension $3^{16} \approx 4.3 \times 10^7$), we perform exact real-time state-vector propagation using a matrix-free implementation of the Hamiltonian.

For larger systems up to $7 \times 7$ ($N = 49$) in 2D, we carry out MPS time evolution via the two-site time-dependent variational principle (TDVP)~\cite{Haegeman2011,Haegeman2016} using \textsc{ITensors.jl}~\cite{Itensor1,Itensor2}. The benchmark MPS simulations in this appendix use a maximum bond dimension $\chi_{\max} = 256$ and a time step $\Delta t = 0.05\,\hbar/J$. Reducing the bond dimension to $\chi_{\max}=128$ changes $\xi^2_{\mathrm{opt}}(t)$ by at most $0.07\,$dB, and $\xi^2_{\min}$ by at most $0.04\,$dB, up to $6\times6$. On the $3\times3$ and $4\times4$ lattices, the MPS and ED minima agree to within $0.013\,$dB.

In Fig.~\ref{fig:SI_benchmark_alpha3_traces}, we compare the full time traces on a $4\times 4$ square lattice with open boundary conditions at $r=0.4$ obtained from ED, MPS, and dTWA. The MPS and ED results for $\xi^2_{\mathrm{opt}}(t)$ [panel (d)] agree to $0.001\,$dB at the squeezing minimum. dTWA captures the qualitative dynamics of the expectation values, variances, covariances, and $\xi^2_{\mathrm{opt}}(t)$,
but the agreement is generally worse than for all-to-all interactions. The deviations mainly reflect slightly slower dynamics in dTWA. %
As a consequence, dTWA underestimates the squeezing and overestimates the optimal squeezing time. 
dTWA reaches its minimum $\xi^2_{\min}=-4.84(4)\,$dB at $Jt_{\rm opt}/\hbar=0.70$, compared with $\xi^2_{\min}=-5.05\,$dB at $Jt_{\rm opt}/\hbar=0.65$ in ED, i.e.\ it underestimates the squeezing by $0.21\,$dB ($\xi^2_{\min}$ about $5\%$ too large) and overestimates $t_{\rm opt}$ by $7\%$. 

Figure~\ref{fig:SI_benchmark_alpha3_scaling} compares the scaling of $\xi^2_{\min}$ with system size up to $7\times 7$ ($N=49$) for all four coupling ratios.
dTWA captures the overall increase of the squeezing with system size, from $\xi^2_{\min}\approx-3\,$dB at $2\times2$ to $-7$ to $-8\,$dB at $7\times7$ in the reference, and the ordering of the curves for $r=0.1$, $0.2$ and $0.4$. For $r\le0.4$, it underestimates the squeezing at every system size, with $\xi^2_{\min}$ from dTWA $2\text{--}15\%$ ($0.08\text{--}0.60\,$dB) larger than the reference. For $r=1$, dTWA underestimates the squeezing for $N\le9$ but overestimates it by $2\text{--}5\%$ ($0.07\text{--}0.20\,$dB) for $N\ge12$.

dTWA captures the optimal squeezing time to within $17\%$ on all lattices. On the lattices accessible to ED ($N\le16$), it overestimates $t_{\rm opt}$ for $r=0.4$ and $r=1$, by up to $14\%$ and $17\%$, respectively, while for $r\le0.2$ the deviations are of either sign and at most $7\%$. On the larger lattices ($N\ge25$), the dTWA $t_{\rm opt}$ for $r\le0.4$ is instead up to $7\%$ earlier than the MPS value, because the reference $t_{\rm opt}$ increases faster with $N$ (from $0.65$ to $0.92\,\hbar/J$ between $4\times4$ and $7\times7$ at $r=0.4$) than the dTWA one ($0.70$ to $0.86\,\hbar/J$). For $r=1$ the two agree to within $5\%$.

\end{document}